**Carbon-based Photocathode Materials for Solar Hydrogen Production**

*Sebastiano Bellani, Maria Rosa Antognazza, and Francesco Bonaccorso**

Dr. S. Bellani, Dr. F. Bonaccorso
Graphene Labs, Istituto Italiano di Tecnologia, via Morego 30, 16163 Genova, Italy
E-mail: francesco.bonaccorso@iit.it

Dr. F. Bonaccorso
BeDimensional Srl, via Albisola 121, 16163, Genova, Italy.

Dr. M. R. Antognazza
Center for Nano Science and Technology @Polimi, Istituto Italiano di Tecnologia, via Pascoli 70/3, 20133 Milano, Italy

## Abstract

Hydrogen is considered a promising environmentally friendly energy carrier for replacing traditional fossil fuels. In this context, photoelectrochemical (PEC) cells effectively convert solar energy directly to $H_2$ fuel by water photoelectrolysis, thereby monolitically combining the functions of both light harvesting and electrolysis. In such devices, photocathodes and photoanodes carry out hydrogen evolution reaction (HER) and oxygen evolution reaction (OER), respectively. Here, we focus on photocathodes for HER, traditionally based on metal oxides, III–V group and II–VI group semiconductors, Si and copper-based chalcogenides as photoactive material. Recently, carbon-based materials have emerged as reliable alternatives to the aforementioned materials. Here, we provide a perspective on carbon-based photocathodes, critically analysing recent research progresses and outlining the major guidelines for the development of efficient and stable photocathode architectures. In particular, we discuss the functional role of charge-selective and protective layers, which enhance both the efficiency and the durability of the photocathodes. We afford an in-depth evaluation of the state-of-the-art fabrication of photocathodes through scalable, high-troughput, cost-effective methods. The most critical issues regarding the recently developed light-trapping nanostructured architectures are also addressed. Finally, we analyse the key challenges on future research directions in terms of the potential performance and manufacturability of photocathodes.

## 1. Introduction

The development of the so-called *Hydrogen economy*,[1] which refers to the vision of using $H_2$ extracted from water through the use of renewable primary sources (wind, solar, geothermal and hydropower) for both energy conversion and storage, ensures near-zero anthropogenic greenhouse gas emissions (*e.g.* $CO_2$, CO, unburned hydrocarbons, $NO_x$), having the same advantages as hydrocarbon fuels, such as a high energy content and usability in conventional power plants.[2] In fact, hydrogen has nearly three times the amount of energy as gasoline (120 MJ $kg^{-1}$ for hydrogen versus 44 MJ $kg^{-1}$ for gasoline),[3] and both internal combustion engines[4] and fuel cells can efficiently use it (*e.g.* the theoretical fuel cell efficiency is above 80%, which is higher than that of gasoline internal combustion engines, lying in the 20 - 30% range)[5]. In this regard, sunlight-powered hydrogen production through photoelectrochemical (PEC) water splitting,[6] referred to as 'artificial photosynthesis', is an attractive solution for tackling fuel demand in the free-fossil era.[7] In an attempt to address this long-standing issue, efficient, long-term stable, cost effective and scalable water splitting PEC devices are needed for an economically competitive hydrogen production on a global energy demand scale.[8] Typically, a water splitting PEC cell comprises a semiconductor photoelectrode immersed in an aqueous electrolyte.[6-8] Semiconductor photoelectrodes absorb light photogenerating electrical charges, which perform the redox chemistry of the hydrogen evolution reaction (HER: $4H^+ + 4e^- \rightarrow 2H_2$) and oxygen evolution reaction (OER: $2H_2O \rightarrow O_2 + 4H^+ + 4e^-$).[8,9] Solar-to-hydrogen conversion efficiency ($\eta_{STH}$) is the most important Figure of Merit (FoM) of a PEC cell, and it is defined by the following equation:

$$\eta_{STH} = \left|\frac{|j_{sc}| \times 1.23 \times \eta_F}{P}\right|_{AM1.5G} \quad (1)$$

in which $J_{sc}$ is the short-circuit photocurrent density, $\eta_F$ is the Faradaic efficiency for hydrogen evolution, and $P_{in}$ is the incident illumination power density, measured under standard solar illumination conditions (AM1.5G).[10] This FoM directly depends on the photophysical properties of the semiconductor photoelectrodes, such as light absorption,[8-10] charge carrier generation/separation[9,10] and transport.[9,10] Moreover, the electrochemical potential value of the photoelectrode conduction band must be below the $H^+/H_2$ redox level ($E_{H+/H2}{}^0$ = 0 V *vs.* standard hydrogen electrode – SHE –),[6,10,11] while the photoelectrode valence band must be higher than the $O_2/H_2O$ redox level ($E_{O2/H2O}{}^0$ = 1.23 V *vs.* SHE).[6,10] Experimentally, tandem PEC cells based on two vertically stacked absorbing materials with different bandgaps can simultaneously optimize the solar light harvesting and increase the photovoltage which, in turn, enhances the photocurrent values.[12] Efficient tandem water splitting systems with $\eta_{STH}$ up to 18% have been demonstrated using III-V compound semiconductors.[12, 13] Very recently, InGaP/GaAs/GaInNAsSb triple-junction solar cell enabled these devices to reach record $\eta_{STH}$ of 30%.[14] However, the high cost (> 2 USD $Wp^{-1}$ for photovoltaic – PV – module) of compound III-V semiconductors is a critical issue.[15] It is noteworthy that $\eta_{STH}$ > 10% has been achieved using cheaper materials than III-V semiconductors[16] such as Si,[17] CIGS[18] and halide perovskites.[19] However, the fabrication of water splitting devices often includes the use of deposition techniques, such as atomic layer deposition (ALD),[20] ion layer adsorption and reaction,[21] and evaporation of metal/metal oxide protective layers[22], which raise the manufacturing costs and/or are not straightforward to scale up. Moreover, the electrolyte-induced degradation of the majority of photoelectrode materials,[13,16] including III-V semiconductors,[13] Si,[17] halide perovskites[19,23] and copper-based chalcogenides,[20a] causes difficulties with regard to the implementation of monolithically integrated devices that are fully immersed in water.[13] However, the most efficient of the aforementioned examples[14,15] are not PEC cells, but rather PV-biased electrosynthetic cells,[24] whose $\eta_{STH}$ is only limited by the electrolysis efficiency (*i.e.* the ratio between the energy content of the $H_2$ and the amount of electricity consumed)[25] of the commercial electrolyzers (up to ~82.3% based on the heating value of $H_2$).[26,27] Unlike PV-biased electrosynthetic cells, in which the light harvesting and electrolysis processes are decoupled (*i.e.* the PV units are externally wired to submerged electrocatalysts),[24] PEC cells with semiconductor photoelectrodes combine both processes via photoelectrolysis[24] which, in principle, could surpass the efficiency of PV-biased electrosynthetic cells.[28] Nevertheless, recent technoeconomic analyses evidenced that PEC water splitting can still not compete with PV-biased electrosynthetic cells,[16,17,28b] which have already been demonstrated and tested in several pilot plants worldwide.[29,30,31]

Therefore, the search for new photoelectrode materials with energy levels compatible with HER,[8,9] electrochemically stable[32] for a long period of time and displaying a low-cost processability into thin film electrodes, is an active research area.[33] For example, current requirements are lifetimes longer than 10 years for an efficiency of 10%, and $H_2$ costs in the range of 2-4 USD $kg^{-1}$ to be competitive with steam-reformed hydrogen;[33a] operational times in the range of 8 years for efficiencies above 3% assuming that the PEC device produces $H_2$ with an amount of energy that is comparable to the energy input required to mine, manufacture and operate the device.[33b]

In this review article, we analyse the challenges, opportunities and potential of carbon-based photocathodes for HER. Photoanode counterparts for OER, photocatalysts, energy-driven water splitting systems[34] (*e.g.* PV-biased electrosynthetic cells)[35] and PV/PEC hybrid systems (*i.e.* PV-biased photoelectrosynthetic cells),[36] which have been the subject of previous reviews,[34,37,38,39,40,41] are beyond the scope of the present paper and are not discussed here.

Based on energy band diagrams, we will first justify the use of carbon-based semiconductors, as promising materials to perform efficiently the PEC HER. Since we focus our attention on the cathodic half-reaction of the water splitting, we will first introduce the FoM that is needed to evaluate the PEC performances of the photocathodes. We will also analyze the latest development of efficient photocathodes, which are progressing at an impressive pace.[42,43,44,45,46,47,48,49,50,51,52,53,54,55,56,57,58,59,60,61,62,63,64] We will also highlight the key-

processes ensuring proper functioning of photocathodes, as well as the main guidelines for achieving efficient and long-lasting device operation. In particular, we will discuss the functional roles of the charge-selective layers (CSLs), electron-conductive layers, and proton conductive membranes that are exploited in the photocathode architectures. We will consider fundamental issues related to the compatibility of photocathode fabrication with a scalable, high-troughput and cost-effective approach, and explore the innovative use of nanostructured scaffolds to enhance the light absorption of the photoactive layer through light-trapping processes. These challenges also present opportunities that must be addressed in order to optimize device design and performance.

## 2. Carbon materials for $H_2$-evolving photocathodes

In the quest for novel photoelectrode materials for HER, carbon-based semiconductors, especially semiconducting polymers (SPs), are emerging as promising candidates due to: their capability to efficiently harvest solar light (their absorption coefficient exceeds $10^5$ $cm^{-1}$)[65]; the tunability of their electronic bandgap and band energies through chemical synthesis (*e.g.* by introducing functional groups and/or heteroatom doping);[66] their charge carrier mobility (up to the order of $10^2$ $cm^2$ $V^{-1}$ $s^{-1}$)[67], which is competitive with amorphous Si (between 0.5-1.0 $cm^2$ $V^{-1}$ $s^{-1}$ for a-Si:H)[68]; their ability to be synthesized in a cost-effective way and be processed in high-throughput productions (in solution-based, roll-to-roll and large-area film depositions)[69]. In fact, the aforementioned properties of the SPs have been already exploited for use in several applications, ranging from organic solar cells (OSCs)[70] and photodetectors/phototransistors[71] to light-emitting diodes[72] and photorefractive devices.[73] The use of carbon semiconductors as photoelectrode materials necessitates three key additional requirements to be fulfilled:[6-12,74] (1) the photoactive material must preserve the capability of generating charges even when it is in direct contact with a liquid electrolyte;[6-12] (2) the photovoltage, which occurs as a result of the optical bandgap and the transport of the photogenerated holes/electrons to the back contact/active surface, must provide sufficient energy to overcome the overpotential that is needed for the electrochemical reactions (*i.e.* HER for the case of photocathodes) at the photoelectrode/electrolyte interface;[74] (3) the kinetics of the interfacial electron transfer at the photocathode/electrolyte interface must be faster than the dynamics of both charge recombination and self-photocorrosion processes.[75]

In this regard, SPs[42,64] and small organic molecules,[76] previously exploited for OSCs,[70] also began to emerge as $H_2$-evolving photocathode materials, since their electrochemical potential of the lowest unoccupied molecular orbital (LUMO) is lower than $E_{H+/H2}{}^0$ (Figure 1a). This is a necessary condition for the production of a photocathode active material.[77] The same also applies to donor-acceptor bulk heterojunction (BHJ) blends established in OSCs,[70, 78] since typical electron acceptors[79] (*e.g.* fullerene derivatives,[78,79] and small molecule acceptors[79,80]) have a LUMO electrochemical potential lower than $E_{H+/H2}{}^0$ (Figure 1b). This allows photocathodes based on the BHJ architecture to be designed.[52-63] Notably, SPs have also been reported as photocathode materials for oxygen reduction reaction (ORR)[81, 82] and hydrogen peroxide ($H_2O_2$) photosynthesis,[83] as well as photoanode materials/photocatalyst for OER,[84,85] thus advancing the design and the realization of all-carbon tandem PEC water splitting cells. Furthermore, several reports[86, 87, 88, 89] have demonstrated the possibility to integrate SPs in different electrochemical devices, including electrochemical sensors,[86] cellular/neuron photostimulation systems,[87] water-gated field effect transistors,[88] and light-powered supercapacitors.[89] Both optical spectroscopy measurements and electrical characterizations proved that they retain the capability to photogenerate electrical charges in aqueous environments.[88b,c,90] The achievement of stable electrochemical behaviour is essential for the development of photocathode technology, in which the instability of the photoactive material when it comes into contact with aqueous electrolyte faces severe issues for practical applications.[22c, 91] For this reason, several photocathode materials have to be protected against direct contact with the electrolyte,[91] resulting in architectures which often resemble PV-biased electrosynthetic cells.[13-19,23] The latter avoids the electrochemical interaction between the photoactive material and the electrolyte, which can result in high-performance photoelectrode designs.[92]

Therefore, by taking advantage of the knowledge that has already been developed in the OSCs field,[67,78] prototypical BHJ-like photocathode structures, which are mainly based on regio-regular poly(3-hexylthiophene-2,5-diyl) (P3HT): with a phenyl-C61-butyric acid methyl ester (PCBM) blend,[49-62] have been used in direct contact with aqueous solution to carry out the HER process (Figure 1c). Recently, this approach has also been extended to graphene derivatives (*e.g.* graphene oxide – GO –[93] and reduced graphene oxide – RGO –[56,93]), which can be used as electrochemically stable interfacial layers that increase the extraction of the photogenerated charges. [56, 93] Notably, producing graphene-based materials from the exfoliation of bulk graphite in suitable liquids[94,95] allows the formulation of functional inks,[96,97] compatible with printing/coating techniques,[98] which are typically exploited for carbon-based photoactive layers.[69] Although they are not the subject of this review, it is also important to highlight that OSC materials, thanks to their aforementioned properties, are also important for the replacement of inorganic PV-driven water splitting technology, which currently achieves $\eta_{STH}$ >18% for AlGaAs/Si tandem solar cells.[13] In fact, tandem and triple junction polymer solar cells have been reported for novel PV-driven water splitting that is compatible with large-area, high-throughput, cost-effective manufacturing.[99] Furthermore, carbon semiconductors such as graphite oxide,[100] GO,[101] carbon dots,[102] graphitic carbon nitride (g-$C_3N_4$)-based materials,[103] conjugated organic microporous frameworks,[104] as well as P3HT,[105] have also been exploited for photocatalytic $H_2$ production, and their $H_2$ production rate is higher than hundreds of μmol $h^{-1}$ for 1 g of an active material under visible light illumination (*i.e.* their quantum yield is > 1% under monochromatic light irradiation).[100,104] However, their use as photocathode materials is almost unexplored. Only recently heterojunction of graphitic carbon nitride/graphdiyne has been reported as photocathode material.[103l]

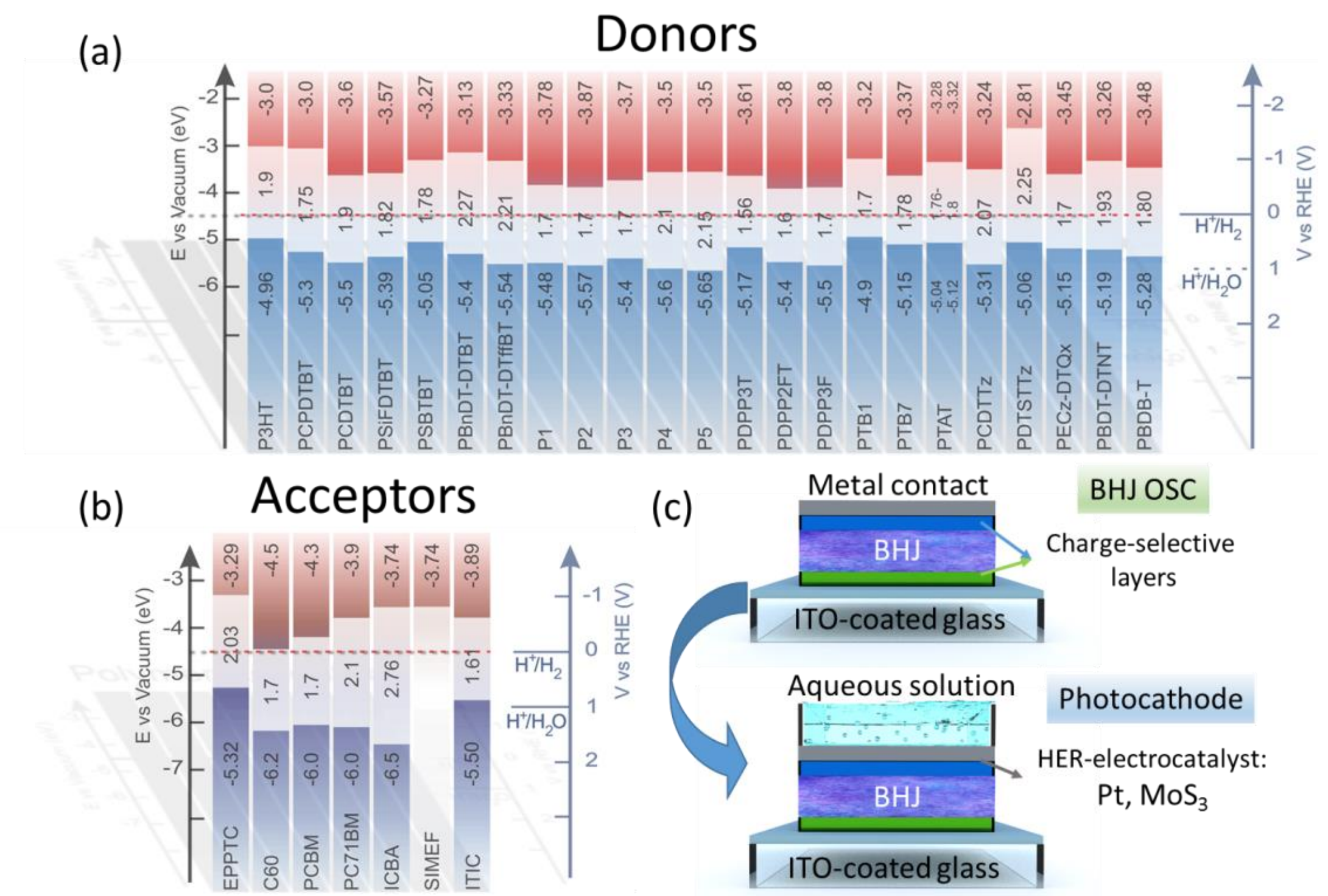

**Figure 1.** Comparison of the highest occupied molecular orbital/lowest unoccupied molecular orbital (HOMO/LUMO) energy (electrochemical potential) levels and band-gap energies of representative a) donor and b) acceptor materials used in OSCs. c) Schematic illustration of BHJ OSCs and the corresponding evolution of the photocathode. The latter is obtained by replacing the metallic cathode with an HER-EC and an aqueous electrolyte.

## 3. Photoelectrochemical characterization and critical parameters

The main FoM that are conventionally used for determining the PEC performance of a photocathode are extracted from the voltammograms, acquired using a three-electrode configuration set-up based on the specific electrolytes (in the form of an acidic, neutral or alkaline aqueous solution). These FoM can be summarized as:[10, 106] (1) the cathodic photocurrent density at 0 V *vs*. the reversible hydrogen electrode (RHE) ($J_{0V\ vs\ RHE}$); (2) the onset potential ($V_o$), defined as ‘the potential at which the photocurrent density relating to the HER reaches a determined threshold value’, which is usually in the order of 10-100 μA $cm^{-2}$; (3) the maximum power point ($V_{mpp}$), which is the potential that satisfies the condition d(JV)/dV = 0; (4) the fill factor (FF), defined as ($J_{mpp}$ x $V_{mmp}$)/($J_{0V\ vs\ RHE}$ x $V_o$), in which $J_{mpp}$ is the current density at $V = V_{mpp}$; (5) the ratiometric power-saved efficiency relative to a non-photoactive (NPA) dark electrode with an identical catalyst (C) ($\Phi_{saved,NPA,C}$), in which identical catalysts are combined with the photoactive material in a co-catalytic configuration; (6) the ratiometric power-saved efficiency relative to an ideally non-polarizable RHE ($\Phi_{saved,ideal}$). $\Phi_{saved,NPA,C}$ is calculated by using the following equation:

$$\Phi_{saved,NPA,C} = \eta_F \times |j_{photo,m}| \times [E_{light}(J_{photo,m}) - E_{dark}(J_{photo,m})] / P_{in} = \eta_F \times |j_{photo,m}| \times V_{photo,m} / P_{in} \quad (2)$$

in which $\eta_F$ is the current-to-hydrogen faradaic efficiency (previously defined in Equation 1), $P_{in}$ is the the incident illumination power density, and $j_{photo,m}$ and $V_{photo,m}$ are the photocurrent density and photovoltage at the $V_{mpp}$, respectively. $j_{photo}$ is obtained by calculating the difference between the current under photocathodic illumination and the current of the corresponding catalyst. The photovoltage $V_{photo}$ is the difference between the potential that is applied to the photocathode under illumination ($E_{light}$) and the potential that is applied to the catalyst electrode ($E_{dark}$) to obtain the same current density. The subscript “m” stands for “maximum”. $\Phi_{saved,NPA,C}$ reflects the photovoltage and photocurrent density of a photocathode independently on the over-potential requirement of the catalyst. $\Phi_{saved,ideal}$ is obtained by considering RHE as a catalyst electrode, *i.e.* setting $E_{dark}$ = 0 V *vs.* RHE in Equation 2. Notably, when the photoactive material is itself the catalyst, $\Phi_{saved,NPA,C}$ is meaningless, and only $\Phi_{saved,ideal}$ can be attained. Stability tests assessing the durability of the photocathodes are carried out by recording the photocurrent density over time in potentiostatic mode (*e.g.* at 0 V *vs.* RHE) under continuous or chopped 1.5AM illumination (*i.e.* simulated sunlight).[10,106]

**4. Evolution of photocathode architectures based on carbon semiconductors**

The evolution of the rational design of the architectures adopted for $H_2$-evolving photocathodes, based on carbon semiconductors as the photoactive component, is sketched in **Figure 2**a.

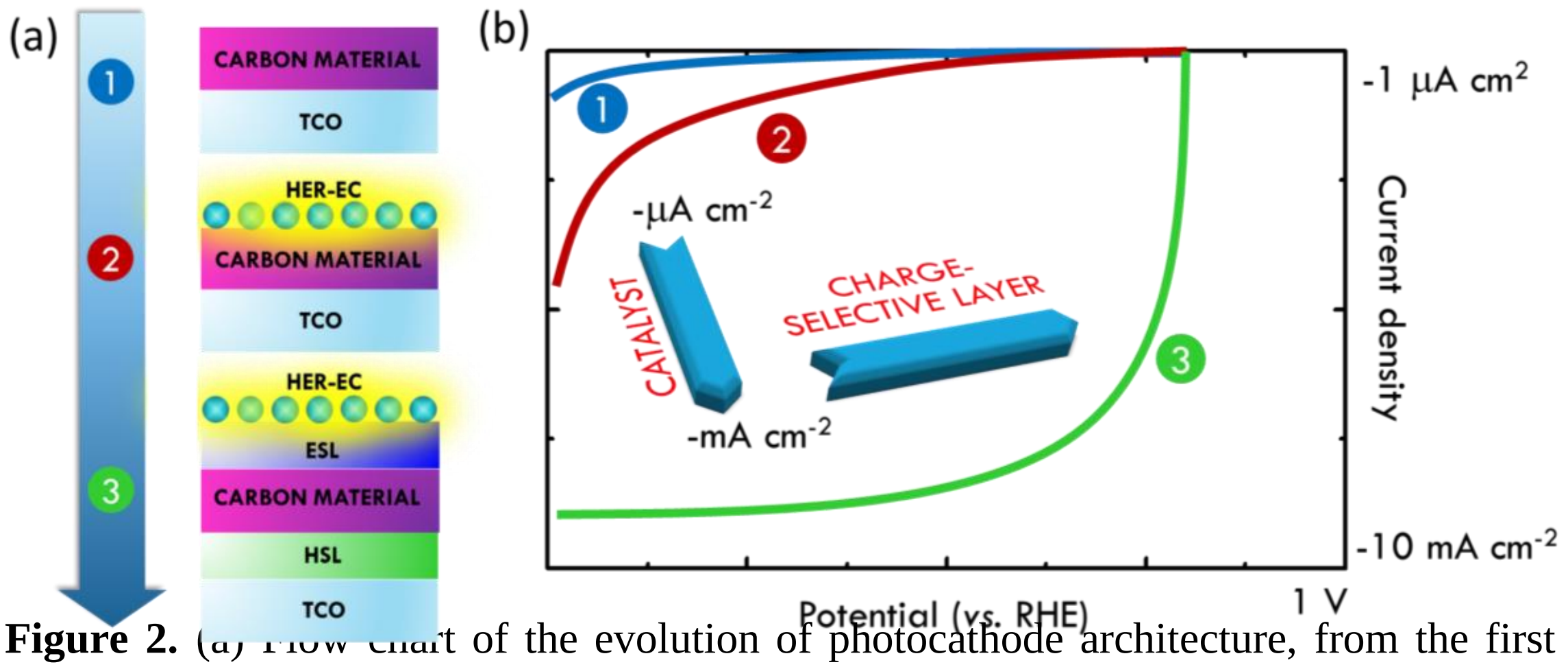

**Figure 2.** (a) Flow chart of the evolution of photocathode architecture, from the first basic design to the most efficient multilayered architectures: (1) carbon material is used as a photoactive material, and as an HER-electrocatalyst (HER-EC); (2) carbon material acts as a photoactive material, while an HER-EC is used as co-electrocatalyst; (3) charge-selective layers (CSLs) are introduced to the design (2), thus creating an OSC-like configuration. (b) Graphical sketch of the PEC behaviour for different architectures.

The first architecture (architecture 1) exploits a carbon semiconductor as photoelectrocatalyst, which both absorbs light and carries out the HER process.[43-46] The second architecture (architecture 2) incorporates an electrocatalyst (EC) for the HER (HER-EC) into the carbon semiconductor, which acts as light absorber.[47] The HER-EC acts as a cocatalyst with the carbon-based materials by lowering the HER activation energy of the latter, which significantly enhances the electrocatalytic activity.[47-49,52-56] The electron transfer between the light absorber and the EC specifically depends on the nature of the materials as well as on their mutual chemical-physical interaction.[107] Therefore, the third architecture (architecture 3) incorporates CSLs in order to improve the electron transfer from the photactive material towards the EC.[50,53,54,56,58,60-62,93]

The incorporation of the HER-EC within the carbon material causes the photocathode photocurrent density to significantly increase (from μA $cm^{-2}$ to mA $cm^{-2}$).[61,62,64,66,68-70,90] In addition, the CSLs raise the HER-operating potential window of architecture (2) to values that are more positive *vs.* RHE.[58,61,62,64,66,68-70,90] Notably, the simultaneous use of an EC and CSLs improves the PEC stability of the light absorber, avoiding photoelectrochemical degradation (*e.g.*, self-photocorrosive reactions).[107] Based on this consideration, the PEC behaviour of the different architectures is graphically sketched in Figure 2b.

### 4.1. Carbon materials as photoelectrocatalysts

As discussed in Section 2, a necessary condition to be fulfilled for the production of a photocathode active material is that it must have a LUMO electrochemical potential lower than $E_{H+/H2}^{0}$ (Figure 1a,b). This requirement is fulfilled by several carbon semiconductors. In particular, P3HT, which is the reference-semiconductive material for OSCs,[108] has a direct bandgap of 1.9 eV[70,77,78] (close to the optimum value for PEC tandem devices), and the LUMO electrochemical potential that is several hundreds of mV lower than the $E_{H+/H2}^{0}$ potential ($LUMO_{P3HT}$ - $E_{H+/H2}^{0}$ < -1 V) (Figure 1a).[70,77,78] Under these conditions, the photogenerated electrons possess the energy that is needed to carry out the HER process.[6,10] Moreover, the optoelectronic properties of P3HT, such as light absorption and charge photogeneration, are still retained in aqueous environments,[87-90] as it has been demonstrated by electrical[88] and optical spectroscopy[90] measurements. Based on this observation, P3HT directly deposited onto transparent conductive oxide (TCO)[109] (*e.g.* Indium Tin Oxide – ITO – ) has been reported as a photoelectrocatalyst for HER.[42,44] The P3HT-photocathodes (architecture 1 in Figure 2a) have demonstrated a photocurrent density of a few tens of μA $cm^{-2}$ in acidic electrolyte (**Figure 3**a).[42,44,45,82,90c] As it has been elucidated in recent studies,[53,79,87c] the photocurrent density is largely due to the reduction of trace molecular $O_2$ (*i.e.* to the ORR: $O_2 + 4H^+ + 4e^- \rightarrow 2H_2O$, for acid condition; $O_2 + 2H_2O + 4e^- \rightarrow 4OH^-$, for basic condition) (Figure 3b).[45,82,90c] The photocurrent density ascribed to HER is instead in the order of sub-μA $cm^{-2}$.[45,82,90c] Spectroscopic and electrochemical measurements revealed the pathways of HER-photoelectrocatalytic activity of the P3HT.[42,44] In more detail, the P3HT surface is protonated through the addition of atomic H at the α-site of the thiophene ring, as sketched in Figure 3c.[44] The protonated P3HT at the polymer/electrolyte interface receives electrons that are photogenerated in the bulk of the P3HT film, resulting in the release of $H_2$ and the reformation of the neutral P3HT surface. The latter is reprotonated for the next reduction cycle, while the photogenerated holes in the film migrate to the back contact to close the circuit at the counter electrode.[44]

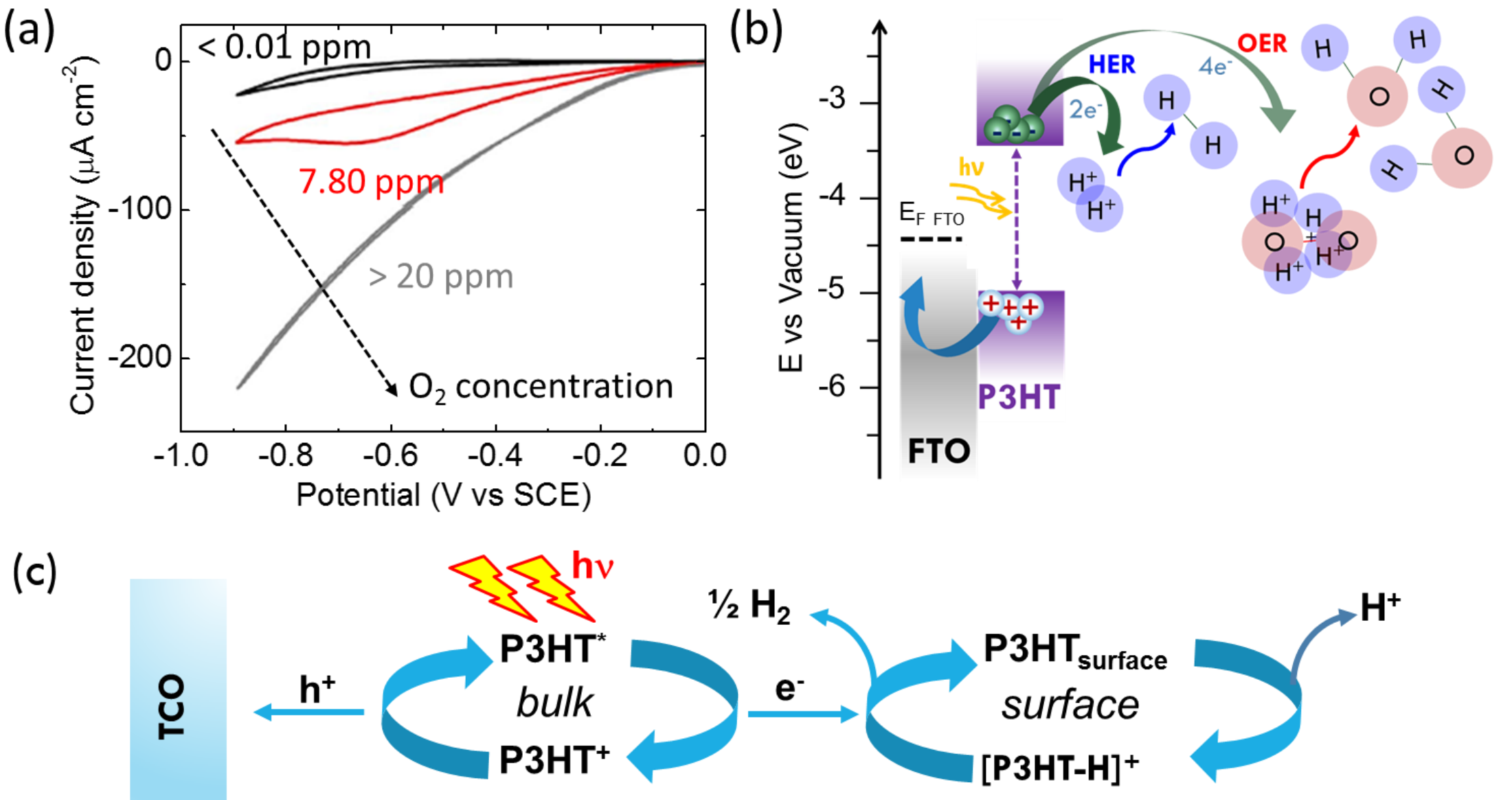

**Figure 3.** (a) Photoelectrochemical response of P3HT coated ITO photocathodes under illumination (100 mW cm$^{-2}$) in 0.1 M $H_2SO_4$ with different dissolved $O_2$ concentrations: <0.01 ppm (black); 7.80 ppm (red); >20.0 ppm (gray). Data reproduced with permission.[45] Copyright 2011, The Electrochemical Society. (b) Scheme of the energy band position of the photocathode materials. The electrochemical reaction pathways, *i.e.* OER and HER, are shown. (c) Proposed mechanism of P3HT HER-activity in aqueous solution.[44]

Similar HER-pathways could also be valid for other SPs, as reported for poly(2,2-bithiophene) (PBTh)-based photocathodes.[46] The photocurrent density can be increased by at least one order of magnitude by mixing SPs and acceptor materials such as ($C_{60}$),[42] metallofullerene[110] and PCBM.[111] This means that, upon illumination, the charge separation mechanism in P3HT:acceptor films immersed in an aqueous solution resembles the one occurring in a BHJ configuration in OSCs. Although these findings encouraged the use of OSC materials for the development of photoelectrode for HER,[43] the observed photocurrent densities (in the order of μA cm$^{-2}$) are two orders of magnitude lower than those expected from the light harvesting efficiency (evaluated from both solid-state OSCs[70,78] and optical absorbance measurements[90]). This observation clearly indicates that the shift in working environments, from solid-state OSCs to solution-based PEC environments, causes severe operative limitations. The main one is ascribed to the sluggish kinetics of the interfacial electron transfer at the solid/aqueous solution interface that leads to the HER process.[45] The low efficiency ($\Phi_{saved,ideal}$ < 0.1 %)[42,43,44] hindered further development in this direction for several years. However, advances in the synthesis and stability of materials,[112] coupled with a deeper knowledge of the main photophysical processes (charge generation,[113,114] transport,[113,115] recombination[113,116]) occurring in OSCs[112-116] and photodetectors,[117] has generated a renewed interest in the exploitation of SPs as photoactive materials in PEC technology. The advances of SPs in the PEC HER field has also been promoted by their development in other applications in aqueous solutions, mainly in the field of electrochemical sensors[86] and biotechnology.[87] These developments have attracted considerable attention, fostering detailed investigations into interface photo-activated mechanisms,[90c] water and oxygen doping,[90a,c] water penetration[90d] and ionic conductivity,[90e, 118] which have contributed to a deep understanding of photocathodes' working processes. As a key achievement in this field, the P3HT:PCBM blend was demonstrated to operate efficiently (*i.e.* the recorded photocurrent density is comparable to that measured in OSCs[70,78]) when the aqueous electrolyte was replaced by an acetonitrile solution containing ferrocene/ferrocene$^+$ (Fc/Fc$^+$) (photoanode configuration) or benzoquinone/benzoquinone$^{\cdot-}$ (BZQ/BZQ$^{\cdot-}$) as the redox couple (photocathode configuration) (Figure 4a,b).[50] These results demonstrated that the capability of the blend to photogenerate the charge is not affected by the interaction with liquid electrolytes (Figure 4c).[50] Consequently, PEC cells that have adopted fluorine doped tin oxide (FTO)/Pt as a counter electrode have photoconversion properties that are comparable to the ones of semitransparent solid-state BHJ OSCs (Figure 4d). Moreover, in the presence of an homogeneous catalyst for HER (cobaloxime, *i.e.*, chloro(pyridine)bis(dimethylclyoximate)cobalt III),[119] replacing the previous redox couples, and a proton source given by a small amount (5 mM) of HCl, the photocathodes reported a photocurrent density of ~ 1 mA cm$^{-2}$ (Figure 4e).[50] Although the experiments were carried out with an organic electrolyte,[50] this example is the first demonstration of carbon semiconductor-based photocathodes achivieng an efficienct HER ($\Phi_{saved,NPA,C} > 1$ %).[50] The experiment reported in Ref. 50 proves that the charges photogenerated in P3HT:PCBM-based PEC devices can be effectively extracted in order to carry out electrochemical reactions, including HER, in a similar way to solid-state OSCs through the use of optimized charge transport layers and metal contacts.[70,78] However, the exclusive use of carbon semiconductors, despite their capability to photogenerate charges, is not enough to efficiently catalyze the HER in aqueous electrolytes, thus evidencing poor HER-electrocatalytic performance.

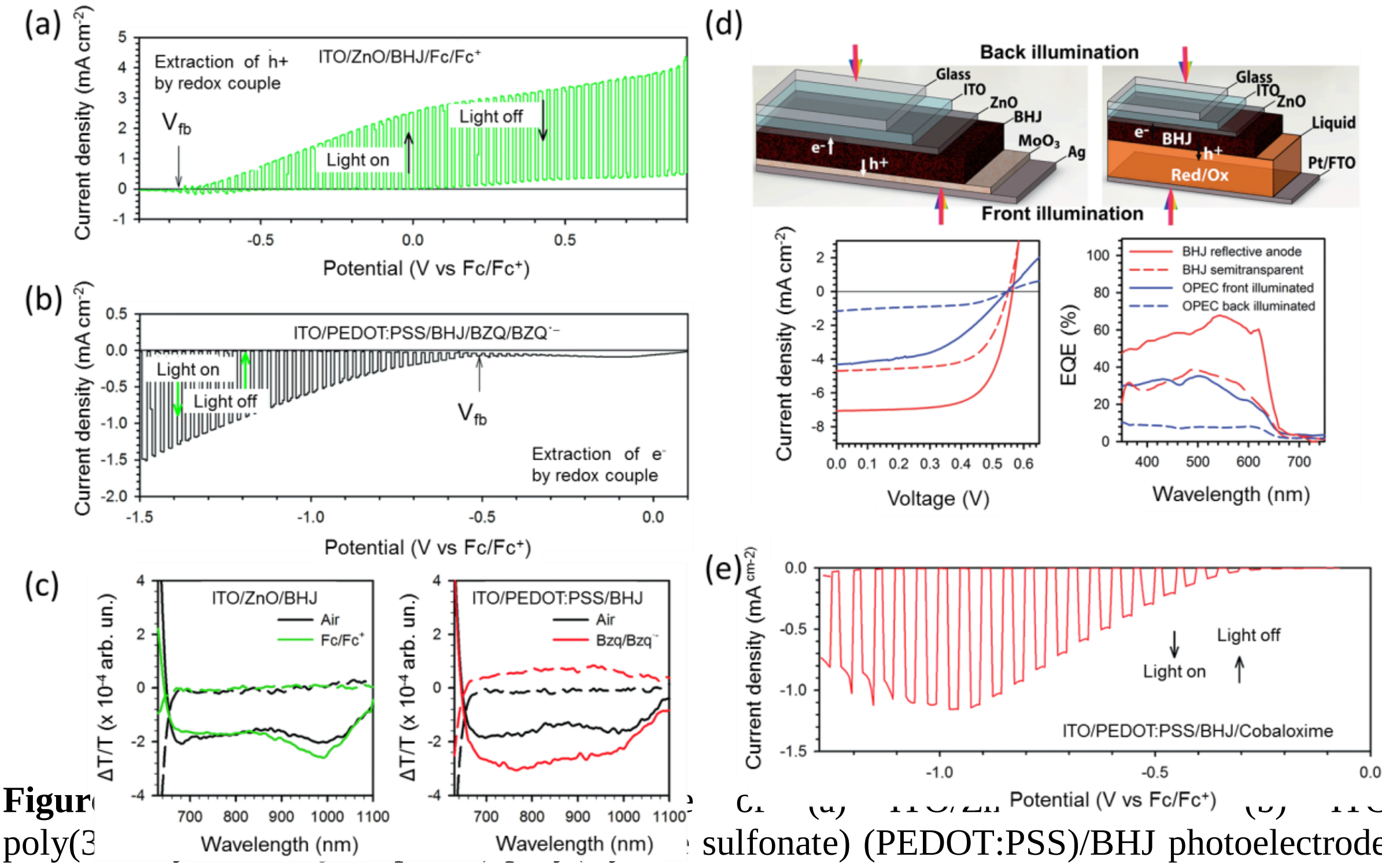

**Figur**... (a) ... O/ poly(3... sulfonate) (PEDOT:PSS)/BHJ photoelectrodes for the redox couple Fc/Fc$^+$ and BZQ/BZQ$^{\cdot-}$, respectively, in acetonitrile (0.1 M tetrabutyl hexafluorophosphate –TBAH) under chopped simulated sunlight (AM1.5 light illumination, 100 W cm$^{-2}$). (c) Continuous-wave photo-induced absorption spectra of (a) ITO/ZnO/BHJ photoanode and (b) ITO/PEDOT:PSS/BHJ photocathode in air and in the presence of Fc/Fc$^+$ and BZQ/BZQ$^{\cdot-}$ redox couples, respectively. Solid lines correspond to the in-phase response and dashed lines to the out-of-phase response. Optical excitation wavelength, 561 nm; modulation frequency, 133 Hz. (d) Comparison between the photoresponse of the BHJ OSC, given by (ITO/ZnO/BHJ/$MoO_3$/Ag, and the PEC cell, given by ITO/ZnO/BHJ/Fc/Fc$^+$/Pt/FTO).

The illumination direction for both devices is indicated in the legend panel. Reflective and semi-transparent BHJ refers to devices with 100 nm and 20 nm thick Ag layer, respectively. Adapted with permission.[50] Copyright 2014, The Royal Society of Chemistry.

### 4.2. Incorporation of HER-electrocatalysts as co-catalysts

The results discussed in Section 4.1 highlight the need for including an EC for HER (HER-EC) in the carbon semiconductors-based photocathode architecture, in order to overcome the poor HER-electrocatalytic activity of the carbon semiconductors. In fact, effective HER-ECs (*e.g.* Pt-group elements,[120] and noble metal-free materials[121] such as transition metal dichalcogenides (TMDs)[122,123,124] and carbides,[125] Co-[126] and Ni-based complexes[127]) have a Gibbs free energy of adsorbed atomic H ($\Delta G_H^0$) close to zero, ensuring a thermodynamically activated HER process.[128] Based on this rationale, research was conducted on modifying the structure of the P3HT and P3HT:PCBM photocathodes.[45,47,52,53,54,56,58] The most common strategy consists in the deposition of the HER-EC (*e.g.*, through sputtering[45] or photoelectrochemical deposition of Pt nanoparticles[47]) on top of the photoactive material (architecture 2 in Figure 2a). In this way, the photocurrent density increased from sub-μA $cm^{-2}$ values for uncatalyzed photoelectrodes to tens of μA $cm^{-2}$ at a positive potential *vs.* RHE, see the graphical sketch in Figure 2b. Experimentally, photocurrent densities in the order of mA $cm^{-2}$ are recorded at a negative potential *vs.* RHE for P3HT:PCBM/Pt photocathodes,[45,47] confirming that the P3HT:PCBM film retains the capability to photogenerate charges in an aqueous solution.

### 4.3. The role of charge selective layers

Despite the beneficial effect on the photocurrent density with regard to the incorporation of HER-EC as a co-catalyst (Section 4.2), the photocathodes that adopted architecture (2) of Figure 2a proved to reliably operate only at a negative potential *vs.* RHE.[45,47,54] This means that the photogenerated charge was not extracted from the photoactive layer efficiently enough to carry out the HER process. Consequently, a limited PEC performance (*i.e.* $\Phi_{saved,NPA,C}$ and $\Phi_{saved,ideal}$ lower than 0.1%) has been experimentally recorded.[45,47,54] In order to enhance the PEC performance of the photocathodes, it is crucial to engineer and optimize the two interfaces photoactive layer/conductive substrate (TCO) and photoactive layer/EC. Actually, the development of interfacial layers has already become crucial with regard to achieving efficient and stable OSCs.[129] The interfaces are based on either insulating, semiconducting or conducting materials. Selective contacts for charge carriers are provided by controlling the material energy levels,[129] as well as the local composition and the phase segregation of the BHJ.[129,130] These features make both OSC concepts and devices directly exploitable in the case of the photocathodes.[49] Architecture (3) in Figure 2a represents this strategy. An ideal selective layer for holes or electrons (HSL and ESL, respectively) has to accomplish the following requirements:[129,131] (1) adjust the energetic barrier height between the active layer and TCO/EC;[112,129-131] (2) transport holes/electrons while blocking charges of the opposite sign;[112,129-131] (3) hinder spurious/parasitic reactions between the active layer and electrode/electrolyte; [112,129-131] (4) ensure a morphology of the active layer suitable for efficient charge dissociation and transport;[129,130] (5) act as an optical spacer,[130] enhancing the light absorption into the photoactive blend by redistributing the optical electric field.[130] Once the electrons are collected in the EC, the HER process is activated thanks to the electrocatalytic activity of the EC itself. Based on these indications, multilayered photocathodes (based on P3HT:PCBM sandwiched between two CSLs)[54,56-58,60,-62,82] enable mA $cm^{-2}$ photocurrent densities at a positive potential *vs.* RHE to be recorded (as schematized in Figure 2b), thus competing with inorganic technologies.[12,18] Experimentally, the HSL is deposited between a TCO as a back conductive contact (*e.g.* ITO, or FTO)[109] and the P3HT:PCBM, while the ESL is deposited onto the P3HT:PCBM. The device is completed by depositing an EC on the HER, resulting in the overall structure TCO/HSL/P3HT:PCBM/ESL/EC.[54,56-58,60,-62,82] As depicted in the scheme of the energy band edge position of the photocathode materials (**Figure 5**), the P3HT:PCBM layer absorbs light and generates charges, while the CSLs maximize the transport of the holes and the electrons toward the TCO and the EC, respectively.

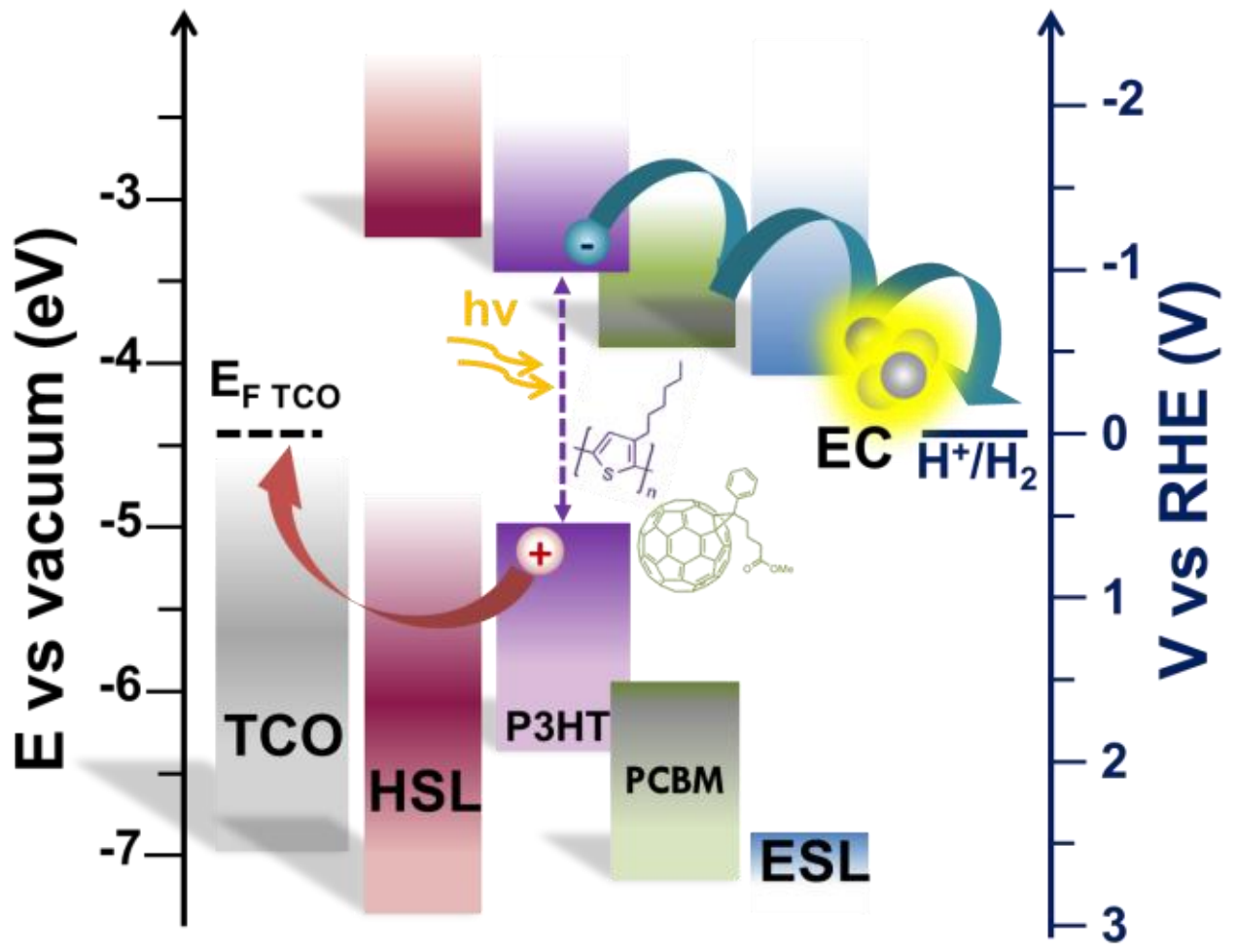

**Figure 5.** Energy band edge positions of photocathode materials, assembled in a multilayered configuration, *i.e.* architecture (3), as reported in Figure 2a. The BHJ layer, typically P3HT:PCBM, absorbs light and generates charges. The HSL drives the holes towards the TCO substrate, while the ESL drives the electrons towards the EC, which perform the HER. The redox level of the HER is also shown.

**Figure 6a** reports a cross-sectional scanning electron microscopy (SEM) image of a representative multilayered photocathode architecture adopting α-$MoO_3$ as HSL, amorphous $TiO_2$ as ESL, and Pt as EC. This is one of the most significant examples of devices structure reported in literature.[54] It is noteworthy that $MoO_3$ is a n-type material, and its apparent hole-selectivity is the result of the formation of a highly p-type doped interface in which SPs have ionization energies lower than the oxide work function.[132] Instead, $TiO_2$ is an established ESL in OSCs,[129,130,133] being also widely implemented in both perovskite and dye-sensitized solar cells as an electron transporting layer.[134] In Ref. 54, $TiO_2$ was deposited onto P3HT:PCBM by pulsed layer deposition (PLD).[135] Actually, PLD is a suitable technique for the deposition of oxide layers on polymeric semiconductors, while still preserving the optoelectronic functionalities of the latter and allowing control over the nanostructure and stoichiometry,[136] thus enabling a large interfacial area with the electrolyte.[54] The effects of the CSLs and Pt, used as an HER-EC, on the photocathode PEC response are shown in Figure 6b. Herein, the photocathode architectures reflect the evolution of the carbon semiconductor-based photocathodes (**Figure 2**). Although the HER performance was promising, the stability of the $MoO_3$-based photocathodes was limited to less than 1 h.[54] This instability was attributed to the irreversible degradation of $MoO_3$ towards sub-stoichiometric phase, such as $MoO_{3-x}(OH)_x$ and $MoO_2$ (Figure 6c,d).[137] Moreover, the energy alignment is not favourable for the role of HSL, since the Fermi energy level is lower than the HOMO energy level of P3HT (*i.e.*, ~-5 eV[70,77,78]).[138]

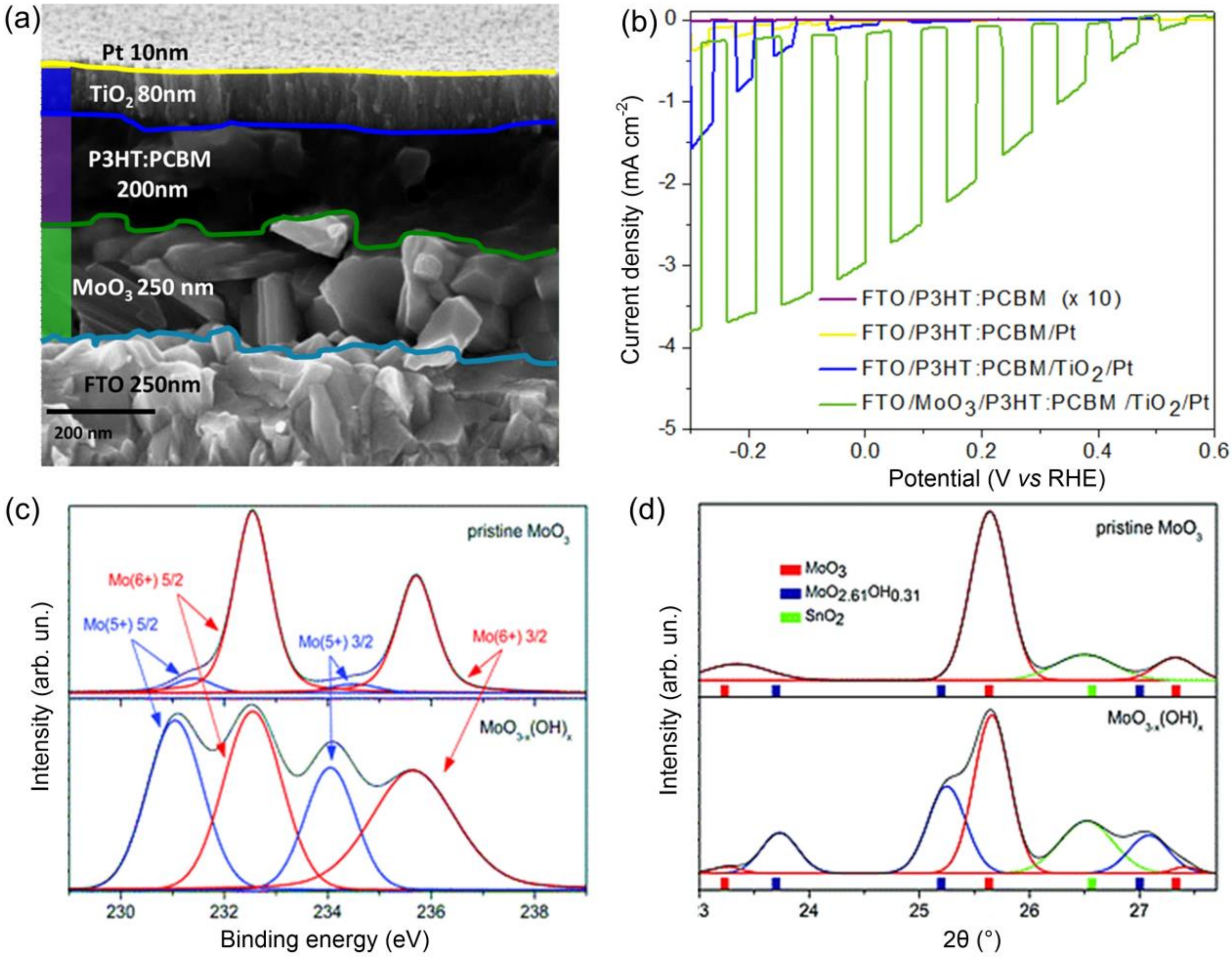

**Figure 6.** (a) Cross sectional SEM image of a representative multilayered photocathode architecture adopting FTO (250 nm) as a transparent conductive substrate, α-$MoO_3$ (~250 nm) as an HSL, P3HT:PCBM (~250 nm) as a light absorber, amorphous $TiO_2$ (~80 nm) as an ESL, and Pt (~10 nm) as an EC. (b) Dependence of the CSLs and the EC on the PEC responses of the photocathodes. The measurements were carried out in a 0.1 M $H_2SO_4$–$Na_2SO_4$ solution,

corrected at pH 1.37, under chopped simulated sunlight (AM1.5 light illumination, 100 W $cm^{-2}$). (c) Mo 3d X-ray photoemission spectroscopy (XPS) and (d) X-ray diffraction (XRD) measurements of FTO/α-$MoO_3$ films before and after polarization in the operative photocathode voltage window. The formation of Mo($5^+$) states and ($MoO_{3-x}(OH)_x$) phase were observed after film polarization. This means that the electrochemical proton intercalation in the $MoO_3$ films occurred while the photocathodes were operating. Adapted with permission from Ref. 54. Copyright 2016, The Royal Society of Chemistry.

It is noteworthy that poly(3,4ethylenedioxythiophene):poly(styrene sulfonate) (PEDOT:PSS), the typical HSL in OSCs, was also tested as an HSL for a P3HT:PCBM photocathode.[52,53] Typically, PEDOT:PSS is heated in air (e.g. at 200 °C for 10 min[53]) to promote the cross-linking to the substrate, *i.e.* to avoid the delamination of the photocathode structures during the operation.[53] Although these architectures were a turning point in the realization of efficient P3HT:PCBM-based photocathodes,[42-44] they still performed poorly (*e.g.* $J_{0V\ vs\ RHE}$ < 1 mA $cm^{-2}$, Figure 7a,b)[52,53] in comparison with the FoM that was expected from the corresponding solid-state P3HT:PCBM-based OSCs ($J_{sc}$ of ~10 mA $cm^{-2}$).[70,108] As in the case of $MoO_3$, the degradation of the photocurrent density (Figure 7c) within the PEC environment has been attributed to the ion penetration and electrochemical doping processes in PEDOT:PSS, which alter the electrical properties of the latter.[139] The presence of $TiO_2$ as an ESL in most of these architectures is not sufficient to protect the underlying structures, as it has been proven by the occurrence of electrochemical reactions involving the HSL being in contact with the electrolytes.[54,57,58,61,63] In fact, despite the photocathode architectures resemble those adopted for OSCs, the devices implemented here are not buried PV cells (*i.e.* PV-biased electrosynthetic system).[24] Importantly, the electrochemical interaction between the photoactive material and the electrolyte is preserved, as denoted by the use of the PEC cell taxonomy.[24] This also enables high-performance PEC designs, such as adaptive semiconductor/EC/electrolyte junctions,[92] to be used even though they are typically excluded for inorganic material-based $H_2$-evolving photocathodes in which the aqueous solutions degrade the photoactive materials.[91]

This discussion focused on some key-examples of PECs[52-54] in which the incorporation of CSLs and HER-EC enhanced the performance of P3HT:PCBM-based photocathodes. However, electrochemical stability issues relating to the CSLs, in particular to the HSLs, caused severe limitations in terms of PEC efficiency and durability. The next section will focus on recent progresses with regard to carbon semiconductor-based photocathodes, providing general guidelines on how to address the PEC stability issue.

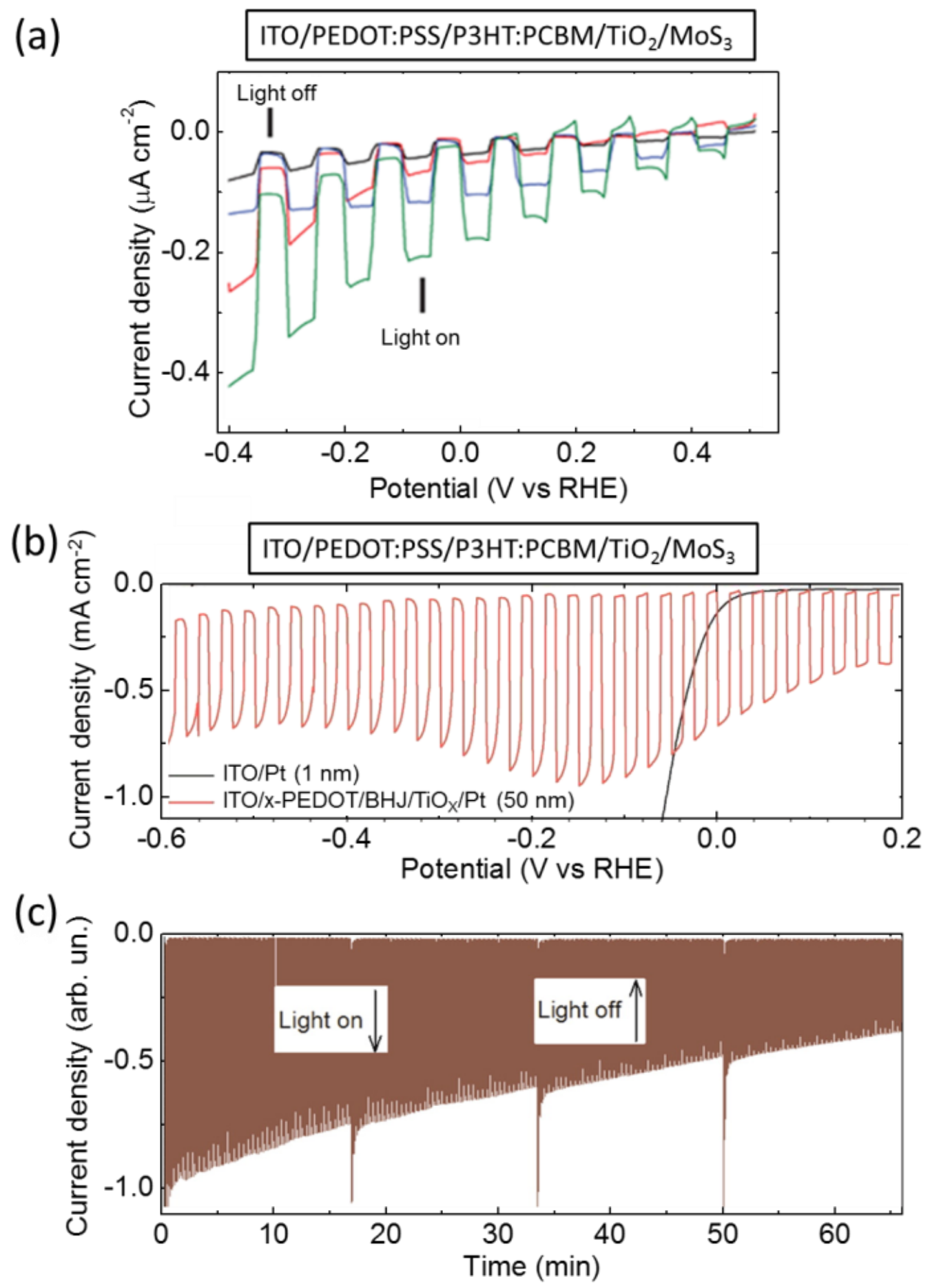

**Figure 7.** Photoelectrochemical response of the first reported multilayered P3HT:PCBM photocathodes adopting PEDOT:PSS as an HSL and $TiO_2$ as an ESL under chopped simulated sunlight (AM1.5 light illumination, 100 W $cm^{-2}$): (a) $MoS_3$ was used as an Earth-

abundant EC and $H_2SO_4$ (0.5 M) as an electrolyte. Adapted with permission.[52] Copyright 2013, The Royal Society of Chemistry. (b) Pt was used as an EC and $H_2SO_4$-$Na_2SO_4$ (pH 2) as an electrolyte. (c) Stability test of the photocathode is shown in (b). Adapted with permission.[53] Copyright 2015, American Chemical Society.

## 5. Strategy to boost the efficiency and long-term stability of carbon semiconductor-based photocathodes

The first stage of the development of P3HT:PCBM-based photocathodes highlighted an important issue, namely the electrochemical degradation of the HSLs that are traditionally used in OSCs, such as PEDOT:PSS[52,53] and $MoO_3$.[54] This had a negative impact on the PEC performance of the photocathodes, both in terms of efficiency (*i.e.* $\Phi_{saved,NPA,C}$ and $\Phi_{saved,ideal}$) and long term stability. In order to overcome this problem, three different strategies have been pursued (**Figure 8**). The first strategy (a) relies on the exploitation of alternative HSL materials (Figure 8, panel a1),[48,56,58,60,61,62,93] including: (1) transition metal oxides (TMOs), *e.g.* NiO[56] and $WO_3$[58]; (2) metal halides, in particular copper iodide (CuI) (panels a2,3)[60,61]; (3) electrically conductive polymers (ECPs), in particular polyaniline (PANI)[48]; (4) two dimensional (2D) materials (*e.g.* graphene-based materials[93] and TMDs, in particular $MoS_2$[62]). The second strategy (b) considers the introduction of a compact electron-conductive interfacial layer between the P3HT:PCBM and the EC (Figure 8, panel b1), which is intended to protect the underlying photocathode structure.[57,63] Different interfacial layers have been tested, including: (1) metallic materials, in particular Ti (Figure 8, panel b2)[57]; (2) high-compact $TiO_2$, as obtained by ALD (Figure 8, panels b3 and b4)[63], which acts both as ESL and protective layer; (3) organic n-type semiconductors, in particular $C_{60}$[57]. Lastly, the third strategy (c) relies on the deposition of a proton conducting overlay onto the photocathode, which prevents the catalyst from dislodging (*e.g.*, catalyst dissolution) and consequent disruptions to the photocathode structure (*e.g.*, delamination effects) (Figure 8, panel c1). To date, two materials have mainly been investigated to improve the photocathode stability: branched polymer polyethyleneimine (PEI)[60,61] and tetrafluoroethyleneperfluoro-3,6-dioxa-4-methyl-7-octenesulfonic acid copolymer (Nafion) mixed with carbon nanoparticles (Figure 8, panels c2 and c3).[93] In the following subsections, we will in depth analyse the aforementioned strategies, discussing the most successful photocathode architectures.

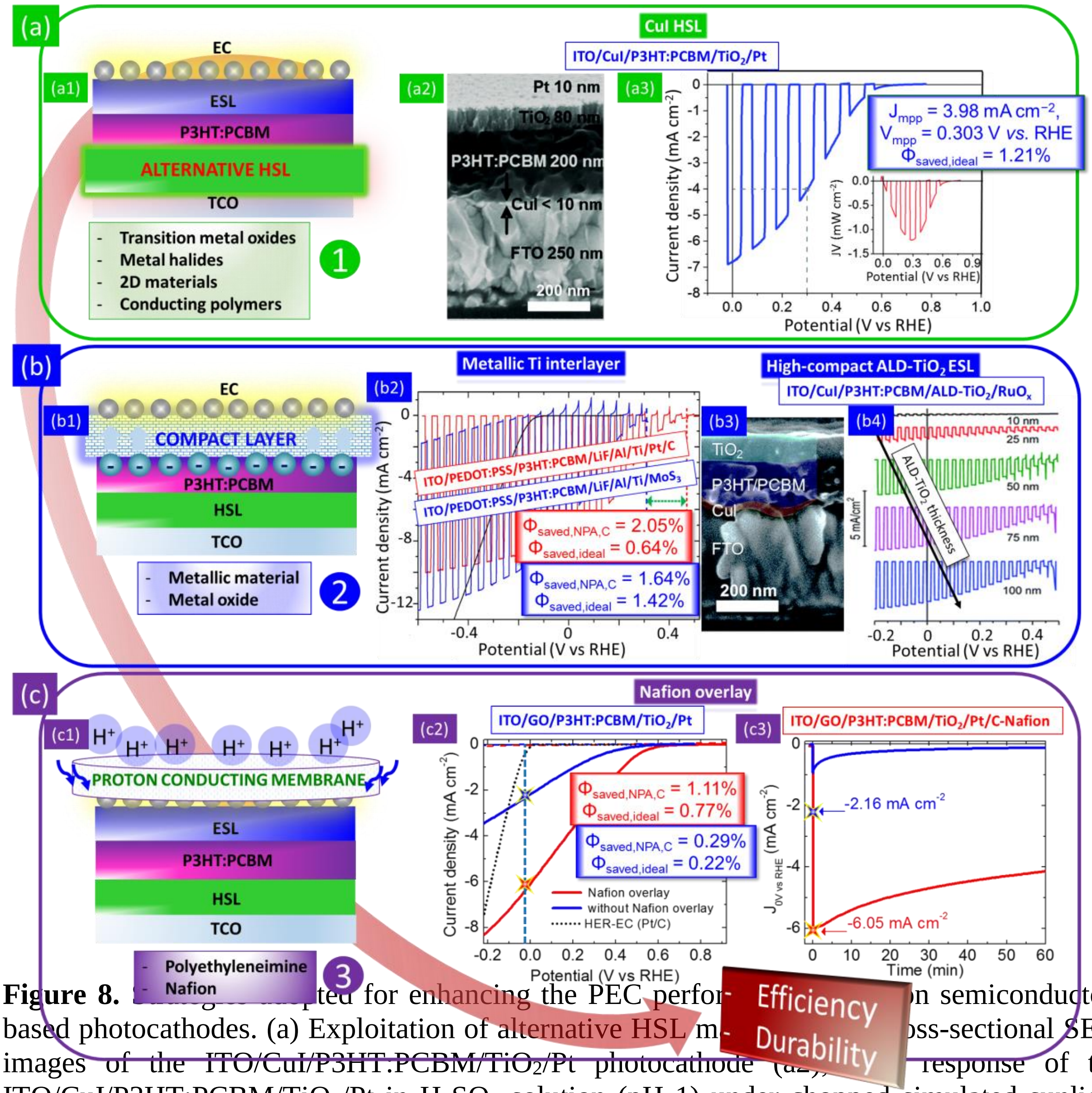

**Figure 8.** [...]ted for enhancing the PEC perfor[...]n semiconductor-based photocathodes. (a) Exploitation of alternative HSL m[...]ss-sectional SEM images of the $ITO/CuI/P3HT:PCBM/TiO_2/Pt$ photocathode (a2), [...] response of the $ITO/CuI/P3HT:PCBM/TiO_2/Pt$ in $H_2SO_4$ solution (pH 1) under chopped simulated sunlight light (a3). Adapted with permission.[60] Copyright 2016, The Royal Society of Chemistry. (b) Introduction of a compact electron-conductive interfacial layer between the P3HT:PCBM and the EC (b1); comparison between the PEC responses of the ITO\PEDOT:PSS\P3HT:PCBM\LiF\Al\Ti\$MoS_3$ (red line) and the ITO\PEDOT:PSS\P3HT:PCBM\Ti\$MoS_3$ (blue line) photocathodes in 0.5 M $H_2SO_4$ under chopped simulated sunlight (b2).Cross-sectional SEM image of a representative $FTO/CuI/P3HT:PCBM/TiO_2/RuO_x$ photocathode, with $TiO_2$ obtained by ALD $TiO_2$ onto the P3HT:PCBM surface (b3); PEC response of the $CuI/P3HT:PCBM/TiO_2/RuO_x$ photocathodes with different $TiO_2$ thicknesses (b4); Adapted with permission.[63] Copyright 2017, The Royal Society of Chemistry. Adapted with permission.[57] Copyright 2015, American Chemical

Society. (c) Deposition of a proton-conducting overlay, which prevents the catalyst from dislodging and prevents consequent structural disruption effects (c1). Comparison between the PEC responses of the FTO/GO/P3HT:PCBM/$TiO_2$/Pt (blue line) and the FTO/GO/P3HT:PCBM/$TiO_2$/Pt/C-Nafion (red line) photocathodes in 0.5 M $H_2SO_4$ under simuated sunlight (c2); stability test of the FTO/GO/P3HT:PCBM/$TiO_2$/Pt (blue line) and the FTO/GO/P3HT:PCBM/$TiO_2$/Pt/C-Nafion (red line) photocathodes (c3). Adapted with permission.[93] Copyright 2017, American Chemical Society.

A summary of the most relevant photocathodes reported in literature, together with their corresponding FoM, is reported in **Table 1**.

**Table 1.** Comparison of the PEC performance ($J_{0V\ vs\ RHE}$, $V_O$, $\Phi_{saved,NPA,C}$ and $\Phi_{saved,ideal}$) of the most relevant carbon semiconductor-based photocathodes that are reported in literature.

| Photocathode architecture | pH | $J_{0V\ vs\ RHE}$ [mA cm$^{-2}$] | $V_O$[a)] [V *vs.* RHE] | $\Phi_{saved,NPA,C}$ [%] | $\Phi_{saved,ideal}$ [%] | Lifetime[d)] (h) | Ref. |
|---|---|---|---|---|---|---|---|
| **Architecture (1) (w/o CSLs and EC)** | | | | | | | |
| ITO/P3HT | 1 | < 0.03 | - | - | - | - | 44,45 |
| Glassy carbon/PBTh | 3, 5 | < 0.1 | - | - | - | - | 48c |
| | 7 | < 0.1 | - | - | - | - | 48 |
| | 8.5, 11 | < 0.1 | - | - | - | - | 48c |
| ITO/P3HT:PCBM | 1 | < 0.1 | - | - | - | - | 47 |
| FTO/P3HT:PCBM | 1.37 | <0.1 | - | - | - | - | 54 |
| FTO/PEDOT:PSS/P3HT:PCBM | 6.8 | 0.045 | 0.55[c)] | | | - | 64 |
| **Architecture (2) (w/ EC)** | | | | | | | |
| ITO/P3HT/Pt | 1 | < 0.1 | - | - | - | - | 45 |
| ITO/P3HT:PCBM/Pt | 1 | < 0.1 | - | - | - | - | 47 |
| FTO/P3HT:PCBM/Pt | 1.37 | <0.1 | - | - | - | - | 64 |
| ITO/P3HT:PCBM/$MoS_3$ | 1 | 0.60 | 0.19 | 0.09 | - | - | 56 |
| FTO/Au/CdS:P3HT/Pt[b)] | 7 | 1.24 | 0.85 | - | - | ~2.5 | 51 |
| **Architecture (3) (w/ CSLs and EC)** | | | | | | | |
| **Screening of HSLs** | | | | | | | |
| ***Conducting polymers (CPs) as HSL*** | | | | | | | |
| FTO/PEDOT:PSS/P3HT:PCBM/Pt | 6.8 | ~0.1 | 0.4 | - | - | - | 64 |
| ITO/PEDOT:PSS/P3HT:PCBM/AZO/C/Pt | 7 | 1.2 | ~0.25 | - | - | ~0.3 | 55 |
| ITO/PEDOT:PSS/P3HT:PCBM/$TiO_x$/Pt | 2 | 0.65 | 0.47 | - | - | - | 53 |
| ITO/PANI/P3HT:PCBM/$TiO_2$/Pt | 1 | 0.3 | < 0.2 | - | - | - | 48 |
| ITO/PEDOT:PSS/P3HT:PCBM/$C_{60}$/$MoS_3$ | 1 | 0.86 | 0.24 | 0.14 | 0.006 | - | 57 |
| FTO/CuI/P3HT:PCBM/$TiO_2$/Pt | 1 | 7.10 | 0.702[c] | - | 1.21 | > 1 | 60 |
| ITO/PEDOT:PSS/α-6T/SubNc/BCP/$C_{60}$/a-$MoS_x$ | 1 | 3.6 | 0.69[d] | - | 0.118 | < 0.5 | 76 |
| ITO/PEDOT:PSS/α-6T/SubNc/SubPc/BCP/$C_{60}$/a-$MoS_x$ | 1 | 1.4 | 0.7[d] | - | 0.027 | < 0.1 | 76 76 |
| ITO/PEDOT:PSS/ α-6T/α-6T:SubNc/SubNc/BCP/$C_{60}$/a-$MoS_x$ | 1 | 2.4 | 0.68[d] | - | 0.087 | < 0.1 | 76 |
| ***Metal oxides*** | | | | | | | |
| FTO/$MoO_3$/P3HT:PCBM/$TiO_2$/Pt | 1.37 | 3.00 | 0.67[c] | - | - | ~ 1.25 | 54 |
| FTO/WO3/P3HT:PCBM/$TiO_2$/Pt | 1.37 | 2.48 | 0.56 | - | 0.246 | ~ 9 | 58 |
| ***Metal halides*** | | | | | | | |
| FTO/CuI/P3HT:PCBM/Pt | - | 1.5 | 0.372 [c] | - | - | - | 60 |
| FTO/CuI/P3HT:PCBM/$TiO_2$/Pt | 1 | 7.10 | 0.702[c] | - | 1.21 | > 1 | 60 |
| **Photocathodes with interfacial compact layer (PV-electrosynthetic cells)** | | | | | | | |
| ITO/PEDOT:PSS/P3HT:PCBM/LiF/Al/Ti/$MoS_3$ | 1 | 8.47 | 0.48 | 2.05 | 0.64 | > 0.17 | 57 |
| ITO/PEDOT:PSS/P3HT:PCBM/LiF/Al/Ti/Pt/C | 1 | 7.87 | 0.67 | 1.64 | 1.18 | - | 57 |
| ITO/PEDOT:PSS/P3HT:PCBM/Ti/$MoS_3$ | 1 | 6.81 | 0.32 | 1.30 | 0.24 | > 0.17 | 57 |
| FTO/CuI/P3HT:PCBM/ALD-$TiO_2$/$RuO_x$ | 1.36 | ~4 | ~0.5 | - | - | > 3 | 63 |
| | 5 | ~2.75 | ~0.5 | - | - | - | 63 |
| **Solution processed-photocathodes** | | | | | | | |
| ITO/PEDOT:PSS/P3HT:PCBM/$TiO_2$:$MoS_3$ | 1 | 0.20 | ~0.3 | - | - | - | 52 |
| FTO/CuI/P3HT:PCBM/$TiO_2$/Pt | 1 | 5.25 | 0.60 | - | 0.63 | ~ 0.5 | 61 |
| FTO/CuI/P3HT:PCBM/$TiO_2$/$MoS_3$ | 1 | 3.94 | 0.57[c] | - | 0.38 | ~ 1 | 61 |
| ***2D material-based HSLs*** | | | | | | | |
| FTO/$MoS_2$/P3HT:PCBM/$TiO_2$/$MoS_3$ | 1 | 1.21 | 0.55 | 0.43 | - | < 0.1 | 62 |
| ITO/RGO/P3HT:PCBM/$MoS_3$ | 1 | 0.70 | 0.35 | 0.11 | - | - | 56 |
| ITO/$NiO_x$/P3HT:PCBM/$MoS_3$ | 1 | 1.3 | 0.34 | 0.24 | | > 1 | 56 |
| ITO/$MoO_x$/P3HT:PCBM/$MoS_3$ | 1 | 2.2 | 0.41 | 0.47 | - | > 1 | 56 |
| ITO/PEDOT:PSS/P3HT:PCBM/$MoS_3$ | 1 | 0.05 | -0.15 | 0.007 | - | - | 56 |
| ITO/RGO/P3HT:PCBM/$TiO_2$/Pt | 1 | 1.33 | 0.50 | 0.18 | 0.15 | < 0.02 | 93 |
| ITO/f-RGO/P3HT:PCBM/$TiO_2$/Pt[e] | 1 | 1.82 | 0.50 | 0.25 | 0.19 | > 1 | 93 |
| ITO/GO/P3HT:PCBM/$TiO_2$/Pt | 1 | 2.16 | 0.50 | 0.29 | 0.21 | - | 93 |
| ITO/f-GO/P3HT:PCBM/$TiO_2$/Pt | 1 | 0.3 | 0.26 | 0.03 | 0.03 | - | 93 |
| **Photocathode w/ protective overlay** | | | | | | | |
| FTO/CuI/P3HT:PCBM/$TiO_2$/Pt/PEI | 1 | 6.8 | 0.715[c)] | - | 1.45 | > 1 | 60 |
| ITO/GO/P3HT:PCBM/$TiO_2$/Pt/C-Nafion[f)] | 1 | 6.01 | 0.60 | 1.11 | 0.77 | > 18 | 93 |
| | 4 | 1.64 | 0.55 | 0.23 | 0.19 | > 2 | 93 |
| | 7 | 1.51 | 0.46 | 0.23 | 0.19 | > 2 | 93 |
| | 10 | 1.41 | 0.60 | 0.23 | 0.20 | > 1 | 93 |
| ITO/f-RGO/P3HT:PCBM/$TiO_2$/Pt/C-Nafion[f)] | 1 | 2.93 | 0.55 | 0.36 | 0.27 | > 20 | 93 |
| | 4 | 0.89 | 0.56 | 0.11 | 0.10 | - | 93 |
| | 7 | 0.91 | 0.54 | 0.12 | 0.10 | - | 93 |
| | 10 | 0.45 | 0.60 | 0.06 | 0.06 | - | 93 |
| **3D nanostructured photocathode** | | | | | | | |
| FTO/ns-MoO3/P3HT:PCBM/$TiO_2$/Pt | 1.37 | ~2 | 0.65[c)] | - | 0.37 | ~ 0.25 | 59 |
| **Large-area photocathode** | | | | | | | |
| 9 $cm^2$-ITO/GO/P3HT:PCBM/$TiO_2$/Pt/C-Nafion[f] | 1 | 2.80 | 0.45 | 0.31 | 0.23 | - | 93 |

[a)] Measured at a photocurrent density of 0.1 mA $cm^{-2}$ (unless specified otherwise) in agreement with the definition of $V_o$ that is given in Section 3. [b)] The P3HT polymer involved in the charge transport process. The role of the light absorber in this device was mostly covered by the CdSe nanoparticles. [c)] Measured when maximum PEC performance has been achieved during the stability test. $V_o$ was 0.55 V *vs*. RHE for the as-produced photocathodes. [c)] Measured at 0.01 mA $cm^{-2}$. [d)] Measured at 0.2 mA $cm^{-2}$. [e)] f-RGO/f-GO are obtained by the chemical functionalization of RGO/GO with (3-mercaptopropyl)trimethoxysilane (MPTMS) in an ethanol solution. [f)] Solution-processed photocathode.[d)] A standard FoM to define the lifetime has not been universally defined in literature. The time at which the photocathodes have shown $J_{0V\ vs\ RHE}$ to be higher than 1 mA $cm^{-2}$ under simulated sunlight is reported (AM 1.5 G light illumination, 100 W $cm^{-2}$).

## 5.1. Exploitation of hole selective layer materials

### *5.1.1. Transition metal oxides*

The search for HSL materials that are electrochemically stable under the working conditions of carbon semiconductor-based photocathodes started with the one already exploited in both organic and inorganic solar cells. Apart the aforementioned $MoO_3$,[54] both NiO[56] and $WO_3$[58] are the most common high work-function TMOs investigated in OSCs, which have been reported as HSLs for P3HT:PCBM-based photocathodes. Pure and stoichiometric NiO is an insulator, while non-stoichiometric $NiO_x$ is a p-type semiconducting oxide,[140] allowing holes to be extracted from its valence band.[56,141] Contrary to $NiO_x$, both $MoO_3$ and $WO_3$ are n-type materials with a conduction band close to the HOMO level of the P3HT.[54,132] As previously mentioned in Section 4.3, the hole-selectivity of $MoO_3$ and $WO_3$ is the result of the formation of a highly p-type doped interface in which SPs have ionization energies lower than the oxide work function, thus allowing the photogenerated holes to be efficiently collected.[132] The solution-processed ITO/HSL/P3HT:PCBM/$MoS_3$ structures that used sol-gel prepared $NiO_x$ or $MoO_x$ as HSLs[56] achieved higher performances compared to both the architecture without HSL and the one exploiting PEDOT:PSS[56] (See Table 1 for values and Ref. 56). In this context, the difference in performance of the different HSLs was attributed to their work function values. The highest onset photovoltage and photocurrent density was achieved by using $MoO_x$, which has the highest work function (*i.e.*, > 5 eV) amongst these three materials.[56] The potential difference between the Fermi levels of the HSL and the electrolyte promotes the migration of holes and electrons towards the TCO and the EC, respectively. This spatially separates the photogenerated charges, thus limiting the occurrence of recombination processes.[56] The $NiO_x$- $MoO_x$- based photocathodes were also more stable than the corresponding architecture without HSL and also the one comprising PEDOT:PSS.[56] In particular, the photocurrent density of the $NiO_x$-based photocathodes initially increased, and after 30 min it stabilized at 1 mA $cm^{-2}$ (see Table 1, Ref. 56). Although the stability tests were limited , i.e., they did not extend longer than 1 h, these results suggested that it is possible to increase the photocathode stability by adopting different metal oxides as HSLs and $MoS_3$ as an EC, and directly depositing them onto P3HT:PCBM.

Following this rationale, amorphous $WO_3$ was studied as an HSL in TCO/HSL/P3HT:PCBM/$TiO_2$/Pt photocathodes[58] by replacing the α-$MoO_3$, previously adopted within an analogue device architecture (see Figure 6).[54] Amorphous $WO_3$ is an n-type semiconductor[141] with a conduction band value close to the HOMO level of the P3HT (~-5 eV[70,77,78]). As is the case with α-$MoO_3$, the hole-selectivity for $WO_3$ is the result of the formation of a highly p-type doped interface in which SPs have ionization energies lower than the oxide work function.[54,132] Although amorphous $WO_3$ has been reported as an electrochemically stable material,[142] reversible proton intercalation/de-intercalation processes have been evidenced,[58] as in the case of α-$MoO_3$. This function is essentially validated by the use of $WO_3$ as a supercapacitor material,[143] thanks to its proton-insertion enhanced pseudocapacitance,[144] and as an additive in the proton exchange membrane[145] due to its proton conductivity.[146] The stability of the $WO_3$-based photocathodes was evaluated at 0.2 V *vs.* RHE, corresponding to the $V_{mpp}$ for the device (Figure 9a).[58] Interestingly, the photocathodes continuously operated for more than 10 h, retaining 70% of their initial photocurrent densities after 8 h, surpassing the durability of previously developed P3HT:PCBM-based photocathodes.[52,53,54,55,56,57] However, the issue of degradation was not completely eliminated, most probably due to the proton intercalation within $WO_3$, a process previously observed in $MoO_3$ HSLs.[54] In fact, proton-intercalated $WO_3$ has a lower work function (~4.5 eV) than that of pristine $WO_3$ (~5 eV).[58] This causes an unfavourable energy level alignment at the $WO_3$/P3HT interface (*i.e.* Fermi level of proton-intercalated $WO_3$ < $HOMO_{P3HT}$).[58] On the contrary, it is important to highlight that proton-intercalation increases the charge carrier density of $WO_3$ by more than two orders of magnitude.[147] This aids the extraction of the photogenerated holes from P3HT to $WO_3$,[54,58,132] counterbalancing the negative effect of the unfavourable energy level alignement at the $WO_3$/P3HT interface.[58] Notably, the multilayered architecture was preserved during the electrochemical intercalation processes (see scanning electron microscopy –SEM– image in Figure 9b), thus accounting for enhanced mechanical integrity of the device.

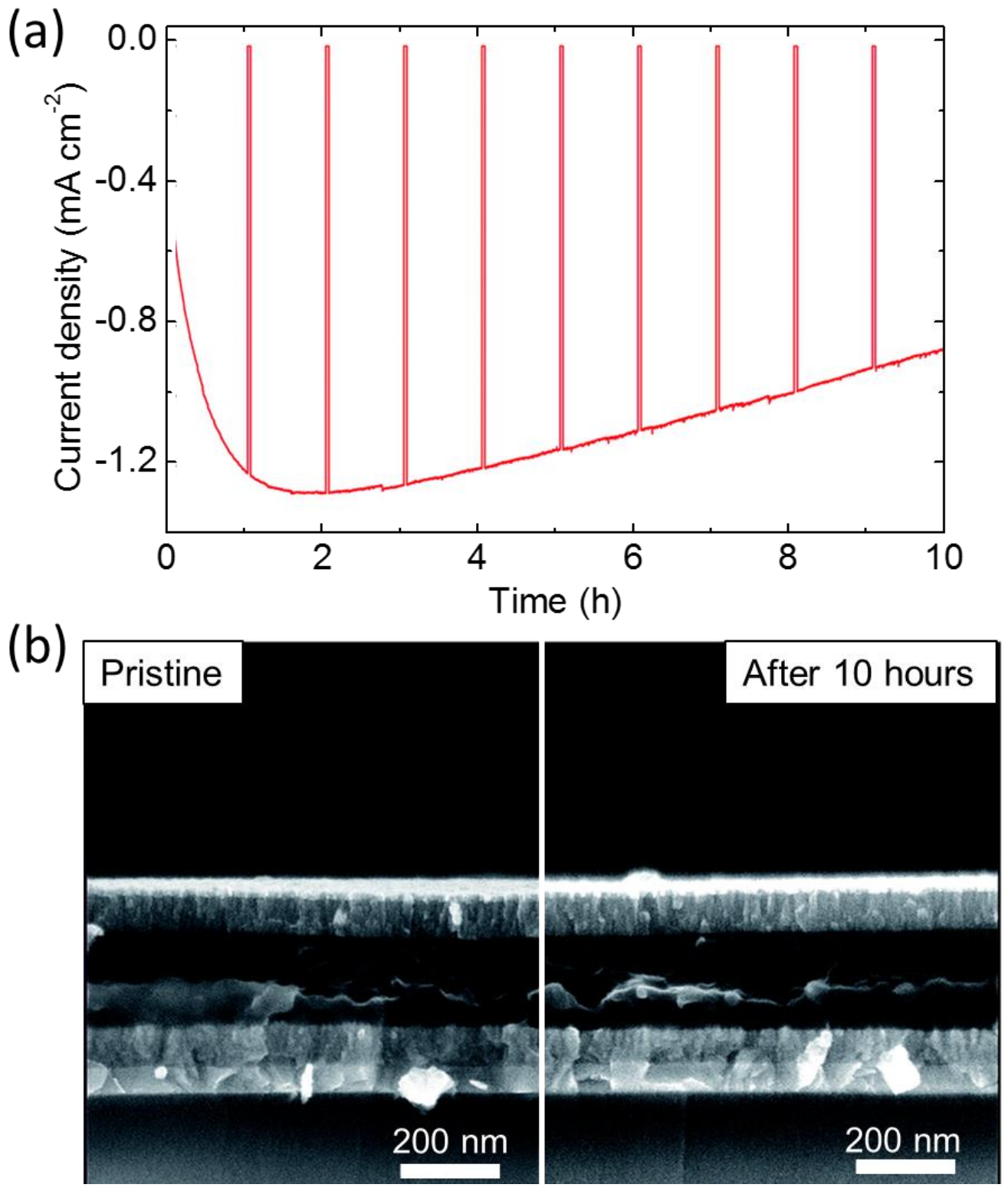

**Figure 9.** (a) Stability test of the ITO/$WO_3$/P3HT:PCBM/$TiO_2$/Pt photocathode under simulated sunlight (AM1.5 light illumination, 100 W $cm^{-2}$) at a fixed potential of 0.2 V *vs.* RHE ($V_{mpp}$). (b) Cross-sectional SEM images of the ITO/$WO_3$/P3HT:PCBM/$TiO_2$/Pt photocathode before and after a 10 h continouos operation at $V_{mpp}$ under simulated sunlight. Adapted with permission.[58] Copyright 2017, The Royal Society of Chemistry.

These results suggest that it is possible to enhance the durability of the photocathode by engineering the electrochemical properties of TMOs as HSLs. Prospectively, *a priori* TMO functionalization (*e.g.* proton intercalation, chemical doping) is a promising strategy for overcoming the photocurrent stability issues.

*5.1.2. Metal halides*

In order to extend the class of HSLs, p-type γ-phase CuI emerged as an efficient low-cost, solution-processable candidate. In particular, the work function of CuI (≥ 4.9 eV)[148] can efficiently extract holes that are photogenerated in the P3HT:PCBM (see Figure 5).[149] Rojas et al. first reported the use of transparent CuI film as an HSL in P3HT:PCBM-based photocathodes through the fabrication of the architecture FTO/CuI/P3HT:PCBM/$TiO_2$/Pt (Figure 8, panel a2).[60] These CuI-based photocathodes exhibited $V_o$ ~0.7 V *vs.* RHE, while the $J_{0V\ vs\ RHE}$ reached a value of 7.10 mA $cm^{-2}$ (Figure 8, panel a3). The $V_{mmp}$ was ~0.3 V *vs.* RHE, corresponding to a cathodic photocurrent density of 3.98 mA $cm^{-2}$. This allowed the $\Phi_{saved,ideal}$ to reach 1.45% (Table 1, Ref. 60). Unfortunately, after 20 min operation, the $J_{0V\ vs\ RHE}$ decreased from ~8.3 mA $cm^{-2}$ to ~3.0 mA $cm^{-2}$. After 1 h, the $J_{0V\ vs\ RHE}$ reached ~1.3 mA $cm^{-2}$. The reason for such performance degradation has been attributed to extended delamination of the Pt layer, as confirmed by scanning electron microscopy analysis (see **Figure 10**a). This observation agreed with previous reports, showing that when Pt is deposited onto $TiO_2$, it is prone to detachment after operating under acidic conditions.[150] In fact, a subsequent second catalyst deposition (replatinization) determined a 73% recovery of the photocurrent density (Figure 10b).[60] The rest of the photocathode structure retained its functionality.[60]

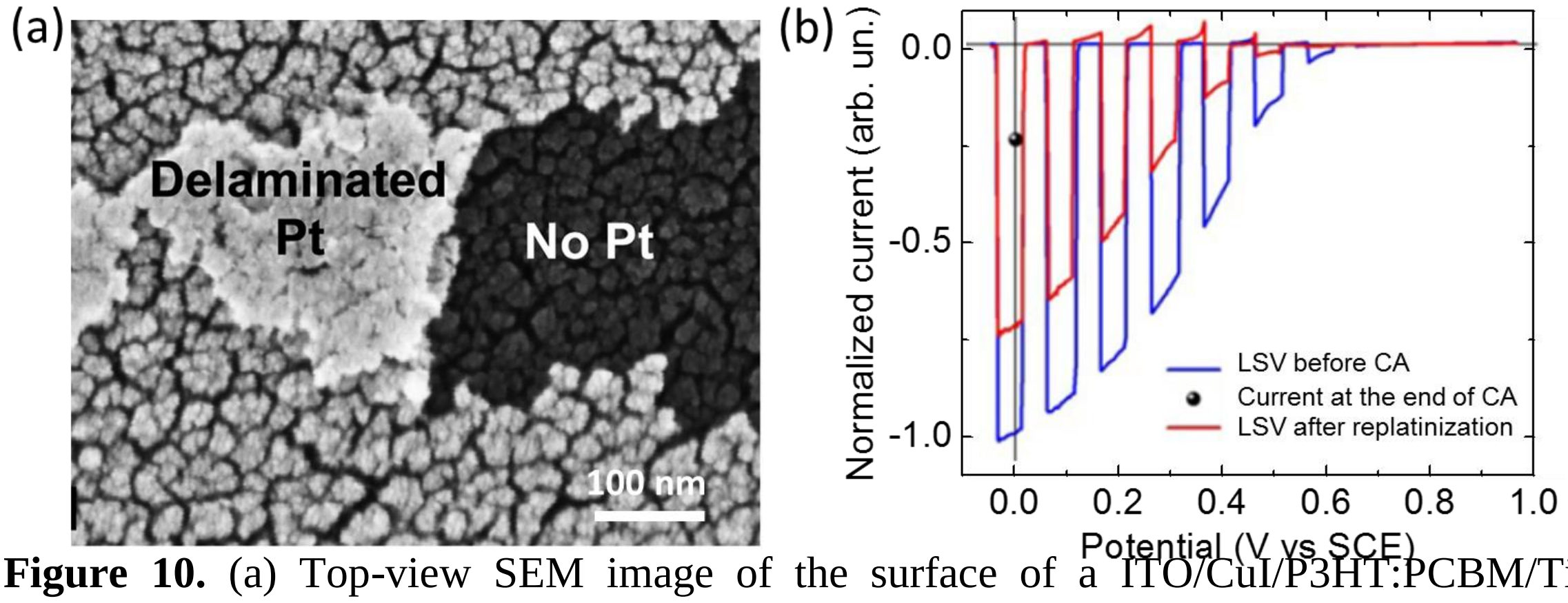

**Figure 10.** (a) Top-view SEM image of the surface of a ITO/CuI/P3HT:PCBM/$TiO_2$/Pt photocathode after 1 h of operation at 0 V *vs.* RHE under simulated sunlight. The image shows the delamination of the Pt layer, with some fragments folding back. (b) Comparison between the PEC response of a ITO/CuI/P3HT:PCBM/$TiO_2$/Pt photocathode after 1 h of operation at 0 V *vs.* RHE, under simulated sunlight, before and after a second Pt deposition (replatinization). The final $J_{0V\ vs\ RHE}$ values after 1 h of operation under simulated sunlight is identified by the black dot. Adapted with permission.[60] Copyright 2016, The Royal Society of Chemistry.

This observation, together with the remarkable PEC performance of the CuI-based photocathode, makes CuI one of the most successful HSL materials to date.[60,61,63]

### *5.1.3. Electrically conductive polymers*

The class of ECPs, such as PANI, Polypyrrole (PPY), and Poly(3,4-ethylenedioxythiophene) (PEDOT), has attracted a great amount of attention over the last two decades due to the inherent electrical conductivity combined with oustanding mechanical/thermal properties.[151] Beyond PEDOT, whose use as HSLs has already been discussed in Section 4.3, PANI is one of the most promising candidates as a p-type semiconductor[152] due to its ease of synthesis on different substrates,[153] the low cost of its monomer (< 10 USD $kg^{-1}$),[154] and its tunable optical,[155] electrochemical[156] and electrical properties.[157] Recently, PANI has been proposed as an HSL in OSCs due to its superior stability compared to other SPs and the tunability of its electrical properties.[129a,158] Based on these previous works, PANI was tested as an HSL in P3HT:PCBM-based photocathodes having the ITO/PANI/P3HT:PCBM/$TiO_2$/Pt architecture.[48] The conductive form of PANI, *i.e.* the emeraldine salt, was obtained from the emeraldine base by doping the polymer with 0.1 M $H_2SO_4$.[48, 159] The PANI-based photocathode has shown a $J_{0V\ vs\ RHE}$ of 0.3 mA $cm^{-2}$ and a photocurrent density higher than 2 mA $cm^{-2}$ at -0.5 V *vs*. RHE.[48] However, the limited value of $V_0$ (~ 0.1 V vs RHE, which was significantly lower than those previously recorded for P3HT:PCBM-based photocathodes (above 0.6 V *vs*. RHE))[54,60,61] and the poor stability of the photocurrent density ($J_{0V\ vs\ RHE}$ loss of 70% after 1 h operation) indicated similar electrochemical issues to those evidenced by using PEDOT:PSS as an HSL (Section 4.3).[53] Although further improvements are surely needed in order to justify the use of PANI as an HSL, these results opened up new perspectives on the use of ECPs as possible HSL materials.

### *5.1.4. 2D materials*

The research of novel HSL materials for P3HT:PCBM-based photocathodes has recently involved 2D materials.[56,62,93] The rationale for the use of 2D materials is linked with the possibility to create and design layered artificial structures with on-demand electrochemical properties by means of large-scale, cost-effective solution processed production methods.[160] In fact, the possibility to produce 2D materials directly from the exfoliation of their bulk counterpart in suitable liquids[94] allows functional inks to be formulated.[96] The latter can then be deposited on different substrates by using established printing/coating techniques.[98] Ref. [62] studied the potential of 2D material interface engineering by using few-layered $MoS_2$ flakes, as a representative TMD, as a HSL in P3HT:PCBM-based photocathodes. Notably, TMDs generated interest due to their optoelectronic properties,[161] for their integration as CSLs into heterojunction-based solar cells, both in OSCs[131a, 162] and in inorganic PV.[163] Moreover, amongst the TMDs, $MoS_2$ is particularly attractive due to its high charge carrier mobility (up to ~470 $cm^2$ $V^{-1}$ $s^{-1}$ for electrons, and ~480 $cm^2$ $V^{-1}$ $s^{-1}$ for holes)[164] and the chemical stability of its basal-planes.[164] In OSCs, solution-processed $MoS_2$ flakes have been exploited as HSLs, and their power conversion efficiency (4% and 8% for P3HT:PCBM and PTB7:PCBM BHJs, respectively)[165] is comparable to that of cells exploiting traditional HSLs, such as $MoO_3$[166] and PEDOT:PSS[167]. More recently, $MoS_2$ has also been exploited in perovskite solar cells [168] either as an hole transporting layer, substituting PEDOT:PSS[169] and Spiro-OMeTAD,[170] or as a conductive and protective buffer layer between the Spiro-OMeTAD and the perovskite layer.[168] Based on these considerations, Ref. [62] investigated a solution-processed architecture, namely FTO/$MoS_2$/P3HT:PCBM/$TiO_2$/$MoS_3$, which adopts both single- and few-layered $MoS_2$ flakes as the HSL, $TiO_2$ as the ESL and $MoS_3$ as the EC. Wet chemical p-doping based on $HAuCl_4$•$3H_2O$ methanol solutions enables the Fermi levels to be tailored.[62] The work function values of the $MoS_2$ films range from 4.6 eV of the pristine $MoS_2$ to 5.1 eV,[62] realizing a favorable energy alignment for the collection of the holes from the HOMO of P3HT (**Figure 11**a, see also scheme of Figure 5).[62] The p-doping process was attributed to the positive reduction potential of the $HAuCl_4$, which accept electrons from $MoS_2$, carrying out the reduction of $Au^{3+}$ to $Au^{0}$ species.[171] Consequently, the as-prepared p-doped $MoS_2$-based photocathodes reached a $J_0$ V*vs*. RHE of 1.21 mA $cm^{-2}$, a $V_o$ of 0.56 V *vs*. RHE and a $\Phi_{saved,NPA,C}$ of 0.43% (Figure 13b), showing a 6.1-fold increase in comparison with the pristine $MoS_2$-based photocathodes.

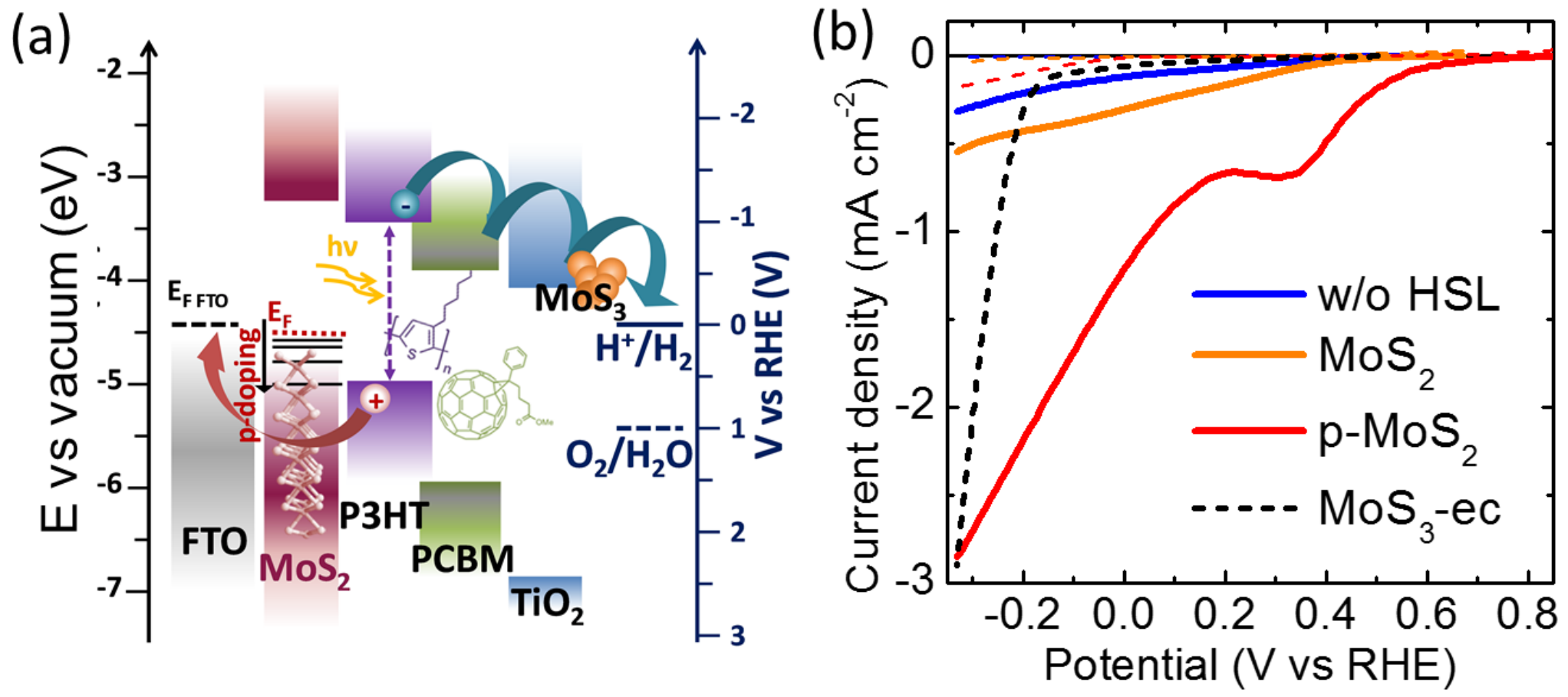

**Figure 11.** (a) Energy band edge positions of the materials that are assembled in the FTO/$MoS_2$/P3HT:PCBM/$TiO_2$/Pt photocathode. The work function values of the $MoS_2$ (4.6 eV) have been measured by an ambient Kelvin Probe. These values can be tailored to higher values (up to 5.1 eV) by wet-chemical doping using $HAuCl_4 \cdot 3H_2O$ as a dopant agent. (b) The photoelectrochemical response of the photocathode with and without $MoS_2$ as HSLs in 0.5 M $H_2SO_4$ solution under simulated sunlight (AM1.5 light illumination, 100 W $cm^{-2}$). Reprinted with permission.[62] Copyright 2017, The Royal Society of Chemistry.

Stability tests have shown an intitial photocurrent density loss, which is attributed to the irreversible $MoS_3$ detachment from the surface of the electrodes, as observed for Pt in similar architectures [54,60] (see Section 5.1.2). This was followed by a progressive stabilization of the photocurrent (photocurrent density loss of 63.2% after 30 min), without any further evidence of delamination of the electrode architecture. Cyclic voltammetry analysis revealed the absence of irreversible redox reactions involving $MoS_2$ films under the operative conditions of the photocathodes, thus suggesting that the FTO/p-$MoS_2$/P3HT:PCBM underlayers were electrochemically stable.[62] These results overall confirmed the electrochemical stability of the $MoS_2$,[172] previously demonstrated in several other energy conversion and storage devices, including fuel and water splitting cells,[173] batteries[174] and supercapacitors.[175] While the implementation of $MoS_2$ is still at the first stages of development, its integration into photocathodes certainly demonstrated its application potential as a stable and efficient HSLs.

Graphene derivatives have also been recently exploited as HSL materials[93], due to their excellent solution processability,[94-98] work function tunability,[176] and good charge transport properties.[94-98,160] For example, GO films, *i.e.*, formed by graphene sheets functionalized with oxygen groups,[98a,177] led to a significant improvement in the performance and stability of the OSCs[178] and perovskite solar cells.[179] The use of solution-processed GO and RGO atomic-thick films as HSLs has recently boosted the efficiency and the durability of P3HT:PCBM-based photocathodes.[93] The photocathodes, based on solution processed FTO/graphene-based HSL/P3HT:PCBM/$TiO_2$/Pt architectures (**Figure 12**a), were fabricated by depositing the material dispersions through subsequent spin coating at low temperature.[93] Experimentally, GO-based photocathodes displayed a better perfomance ($J_{0V\ vs\ RHE}$ = -2.16 mA $cm^{-2}$, $V_o$ = 0.56 V vs. RHE, $\Phi_{saved,NPA,C}$ = 0.29%, $\Phi_{saved,ideal}$ = 0.21%) than RGO ($J_{0V\ vs\ RHE}$ = -1.33 mA $cm^{-2}$, $V_o$ = 0.50 V vs. RHE, $\Phi_{saved,NPA,C}$ = 0.18%, $\Phi_{saved,ideal}$ = 0.15%) (Figure 12b, see also Table 1, Ref. 93). However, despite the promising FoM values, the photocathodes have shown $J_{0V\ vs\ RHE}$ losses of ~95% and ~93% for GO- and RGO-based ones, respectively, after 1 h operation.[93] Performance degradation was attributed to both the detachment/dissolution of Pt from the $TiO_2$ surface[150] (as previously reported for photocathodes under acid conditions),[54,60] and to the poor adhesion between the different layers of the FTO/GO(RGO)/P3HT:PCBM structure after being immersed in the electrolyte.[180] The latter issue was partially overcome by the silane-based chemical functionalization of GO and RGO, which allowed hydrogen-bonded FTO/graphene-based HSL/ P3HT:PCBM structures to be fabricated with improved mechanical adhesion properties (Figure 12c).[181] As reported in Figure 12d, the PEC performance drastically decreased for the photocathodes based on f-GO ($J_{0V\ vs\ RHE}$ = -0.30 mA $cm^{-2}$, $V_o$ = 0.23 V *vs.* RHE and $\Phi_{saved,ideal}$ = 0.03%) in comparison with those recorded for the photocathode using GO (Figure 12b).[93] However, a clear enhancement of the performance was observed for photocathodes based on f-RGO ($J_{0V\ vs\ RHE}$ = -1.82 mA $cm^{-2}$, $V_o$ = 0.5 V *vs.* RHE and $\Phi_{saved,ideal}$ = 0.19%) if compared with those obtained for RGO-based photocathodes (Figure 12b).[93] In fact, the presence of silane groups altered the dipole formation between f-GO and P3HT:PCBM,[182] thus the extraction of holes through quantum mechanical tunnelling processes were negatively affected.[131a,183] However, the functionalization of the RGO flakes, which extract the charge carriers directly through their valence band,[131a,178c,183] enabled more homogeneous film deposition, thus improving the quality of the contact between FTO/HTL and P3HT.[131a,178c] After 1 h operation, the f-RGO-based photocathode still provided a $J_{0V\ vs\ RHE}$ of ~1 mA $cm^{-2}$, and there was no evidence of delamination/disruption of the photocathode structure.[93] These results demonstrated that the functionalization of RGO flakes is an effective tool to strengthen the adhesion between the layers of the FTO/HSL/rr-P3HT:PCBM structure, thus increasing both the efficiency and the stability of the photocathodes. However, it is worth noting that the photocathodes using f-RGO still show degradation effects, such as the detachment of the Pt layer. [54,60] This is an indication that there are other causes of instability for the designed photocathodes.[54,60] The implementation of a protective overlay in order to overcome the latter issue will be discussed in Section 5.3, with the focus on the results achieved for the photocathode exploiting CuI[60,61] and the use of graphene derivatives as HSLs.[93]

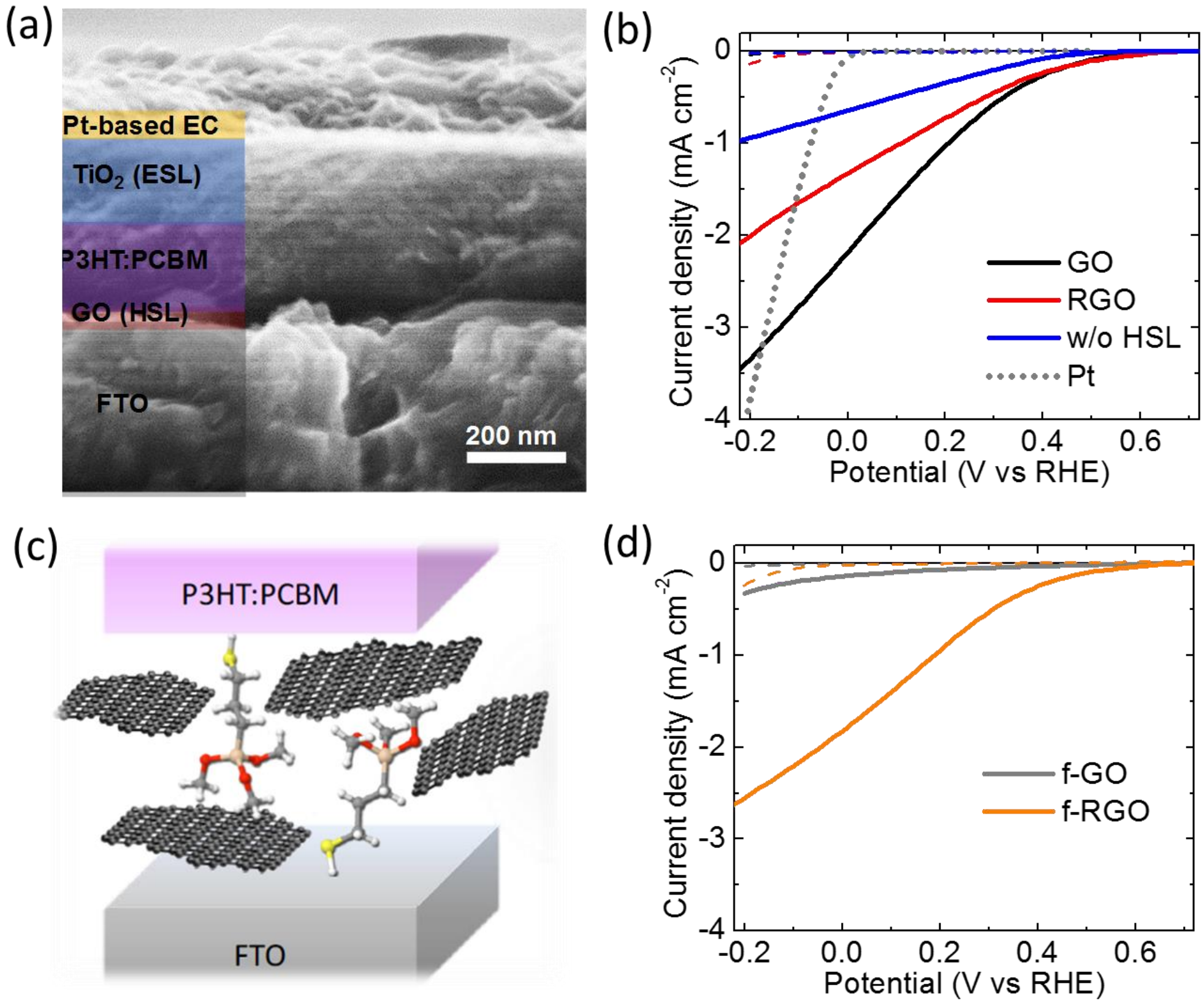

**Figure 12.** (a) Cross sectional SEM image of a representative FTO/GO/P3HT:PCBM/$TiO_2$/Pt photocathode. (b) Photoelectrochemical response of the FTO/graphene-based HSL/P3HT:PCBM/$TiO_2$/Pt photocathodes in 0.5 M $H_2SO_4$ solution under simulated sunlight (AM1.5 light illumination, 100 W $cm^{-2}$). (c) Silane-based chemical functionalization of GO(RGO) for the fabrication of hydrogen-bonded FTO/graphene-based HSL/rr-P3HT:PCBM structures. (d) Photoelectrochemical response of the FTO/graphene-based HSL/P3HT:PCBM/$TiO_2$/Pt photocathodes in 0.5 M $H_2SO_4$ solution under simulated sunlight (AM1.5 light illumination, 100 W $cm^{-2}$), after the silane-based chemical functionalization of the HSLs (f-GO and F-RGO). Reprinted with permission.[93] Copyright 2017, American Chemical Society.

### 5.2. Interfacial compact layers

In Section 4.3, we evidenced that the use of PEDOT:PSS as an HSL raised concerns about electrochemical stability.[52,53,57] In fact, ion penetration 184 and electrochemical doping processes[139] alter the electrical properties of pristine PEDOT:PSS, thus negatively affecting the PEC perfomance of the photocathodes.[52,53,57] In order to mantain the native electrochemical properties of PEDOT:PSS, compact and electron conductive layers between the P3HT:PCBM layer and the EC have been considered (Figure 8, panel b1).[57] Ideally, such interfacial layers should: (1) provide an electron-selective contact with the photoactive layer, reducing the recombination losses and enhancing the $V_o$;[50,54,57,63] (2) form an electron-conductive channel between the photoactive layer and the EC;[50,54,57,63] (3) prevent water penetration with a pinhole-free barrier, thus reducing the electrochemical degradation of the underlying materials.[57,63] A key requirement of the interfacial layer is the electrochemical stability under the operative conditions of the photocathodes.[57,63 Having considered these issues, ITO/PEDOT:PSS/P3HT:PCBM/LiF/Al/Ti/EC architectures[57] based on PEDOT:PSS as an HSL and LiF/Al as an ESL (which are typically used for cathodes in OSCs)[70,78] exploited a metallic Ti layer, evaporated onto Al, to protect the underlying photocathode structure.[52] Inorganic material-based photocathodes exploited metallic Ti as a protective layer in a similar way.[22c] As well, in the case of P3HT:PCBM-based photocathodes the use of a protective Ti layer enabled a significant improvement of the PEC performance (Figure 7a,b). In particular, a $J_{0V\ vs\ RHE}$ of 8 mA cm$^{-2}$ and a $V_o$ value of ~0.48 V *vs*. RHE were measured for ITO/PEDOT:PSS/P3HT:PCBM/LiF/Al/Ti/$MoS_3$ photocathodes (Figure 8b2 and **Figure 13**a. See also Table 1, Ref. 57). The onset of the dark current density was at -0.15 V *vs*. RHE, which is in agreement with the HER-overpotential of $MoS_3$.[185] Thus, the light-driven anodic shift of the HER onset potential (*i.e*. the difference between the $V_o$ and the onset of the dark current density) was ~0.63 V, resembling the open-circuit voltage of the OSC (Figure 13a).[57] Remarkably, the PEC response (a $J_{0V\ vs\ RHE}$ of 7.87 mA cm$^{-2}$ and a $V_0$ of 0.67 V *vs*. RHE) was obtained by replacing $MoS_3$ with Pt/C (Figure 13b).[57] The difference in the HER-overpotential between $MoS_3$ and Pt ECs reflected the difference in the $V_o$ that was observed in the two photocathodes.[186] Despite the photocathodes reached a PEC performance comparable to the one expected from the corresponding solid-state OSC (the short circuit current is in the order of 10 mA cm$^{-2}$ and it has an open circuit potential of ~0.6 V),[57] the photocurrent density decreased during the stability test (*i.e*., $J_{0V\ vs\ RHE}$ reduction of 45% after 10 min).[57] This instability was caused by the lift-off of the LiF/Al/Ti:$MoS_3$ metallic layer when the electrolyte reached the LiF/Al layer. To avoid this degradation process, photocathodes were fabricated without the LiF/Al layer. However, the $V_o$ of these new photocathodes was 150 mV more negative than the value obtained by adopting the LiF/Al layer (see Table 1, Ref. 57). The decrease in the $V_o$ was attributed to the changes in the electron injection barrier at the interface between P3HT:PCBM and the metallic layer,[187] as expected based on the different work functions of the metals.[188] The photocathode using Ti as the sole interfacial layer had a $J_{0V\ vs\ RHE}$ loss of only 12% after 10 min, which enhanced the stability of the photocathode adopting the LiF/Al/Ti layer. Moreover, the Ti layer did not peel off during the 1 hour-long stability test. Overall, the use of the Ti layer preserved the HSL-role of the PEDOT:PSS, increasing the photocurrent density from ~hundreds of μA $cm^{-2}$ (Figure 7a) to ~10 mA $cm^{-2}$ (Figure 13a,b). These results prove that metallic Ti layers limit the contact between the underlying photoactive structure and the electrolyte, which is beneficial for the electrochemical stability of both HSLs and ESLs.

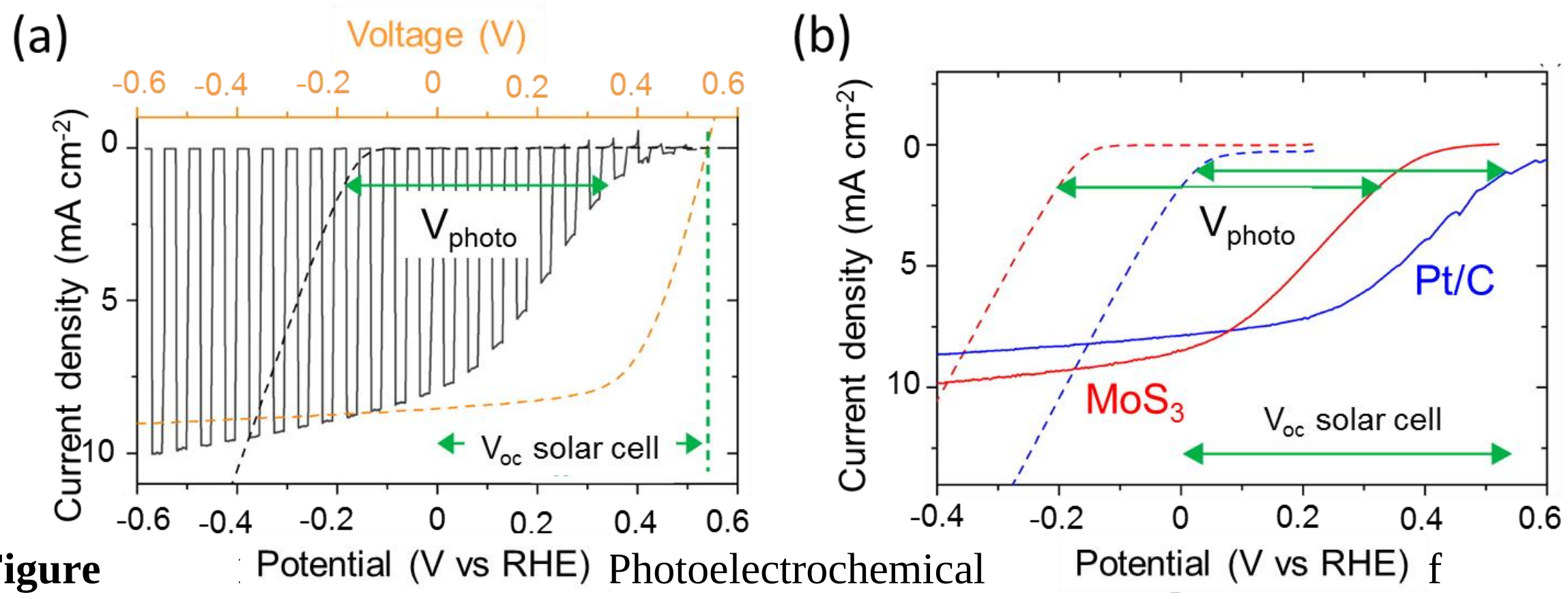

**Figure** Photoelectrochemical f the ITO/PEDOT:PSS/P3HT:PCBM/LiF/Al/Ti/$MoS_3$ photocathode in 0.5 M $H_2SO_4$ under chopped simulated sunlight (AM1.5 light illumination, 100 W $cm^{-2}$) (black solid line, bottom axis). The HER-activity of the ITO\$MoS_3$ cathode in 0.5 M $H_2SO_4$ is also shown (black dashed line, bottom axis). The J-V curve of an ITO/PEDOT:PSS/P3HT:PCBM/LiF/Al OSC is shown for comparison (orange dashed line, top axis). (b) Comparison between the PEC response of the ITO/PEDOT:PSS/P3HT:PCBM/LiF/Al/Ti/$MoS_3$ (red line) and the ITO/PEDOT:PSS/P3HT:PCBM/LiF/Al/Ti/Pt/C (blue line) photocathodes in 0.5 M $H_2SO_4$ under simuated sunlight. The HER-activities of the ITO\$MoS_3$ cathode (red dashed line) and an ITO/Pt/C cathode (blue dashed line) are also shown.
Adapted with permission.[57] Copyright 2015, American Chemical Society.

Evaporated $C_{60}$ (50 nm), an n-type fully organic semiconducting layer used in OSCs,[70,78] was also tested as an alternative material to the metallic Ti layer as interfacial layer between Al and EC.[57] The hydrophobic nature of $C_{60}$[189] was expected to ensure the photocathode stability by preventing the wetting of the PEDOT:PSS/P3HT:PCBM structure underneath.[139] The PEC performance of these photocathodes ($J_{0V\ vs\ RHE}$ near to 1 mA $cm^{-2}$) outperformed those obtained without an interfacial layer.[57] However, the recorded photocurrent densities were remarkably lower than those measured for both the corresponding ITO/PEDOT:PSS/P3HT:PCBM/LiF/Al solid-state OSC and the ITO/PEDOT:PSS/P3HT:PCBM/LiF/Al/Ti:$MoS_3$ photocathode.[57] Moreover, the photocathodes based on $C_{60}$ rapidly degraded during the acquisition of different PEC responses.[57] The instability of the photocathodes was ascribed to the water diffusion through the $C_{60}$ layer towards the PEDOT:PSS/P3HT:PCBM structure underneath. Despite these results, the idea that other fullerene derivatives can potentially act as better interfacial protective layers than $C_{60}$ is still valid.

As an alternative approach to the use of both metallic and organic interfacial layers, low-temperature ALD of high-compact $TiO_2$ onto a P3HT:PCBM polymer blend surface has recently been investigated (Figure 8, panels b3 and b4).[63] The protective action of compact $TiO_2$ layers has been successfully demonstrated on a variety of corrosion-sensitive photocathode materials.[22c,19a, 190] Typically, corrosion protection requires conformal and pinhole-free $TiO_2$ film, which is difficult to achieve by using solution- or physical vapor-based deposition methods.[22c,19a,63,191] Atomic layer deposition allows for growing continuous, conformal metal oxide films with nanometer-scale thickness control over large deposition areas[192] through alternating the exposure of vapor-phase metal-organic and oxidant precursors.[192, 193] The ALD method is commonly used in the fabrication of inorganic semiconductor devices for microelectronic applications,[194] wherein the substrate surfaces are metals or metal oxides.[194,195] Conversely, its application in organic substrates, such as SPs, has scarcely been explored.[192,196] In fact, organic materials may decompose at temperatures typically used in thermal ALD processes, and oxidation processes may occur during oxidant exposure.[63] Despite these challenges, the ALD of $TiO_2$ onto P3HT:PCBM surfaces has been recently achieved by controlling the deposition temperature (80 °C), in order to avoid polymer degradation, using tetrakis(dimethylamido)titanium (TDMAT) as a reactive Ti precursor, and water as a mild oxidant.[63] A continuous $TiO_2$ film atop the polymer blend was formed by using the ALD process, without negatively affecting the blend photoactivity.[63] In fact, FTO/CuI/P3HT:PCBM/ALD-$TiO_2$/$RuO_x$ photocathodes (see cross-sectional SEM image in Figure 8b3) reached a $J_{0V\ vs\ RHE}$ of 4 mA $cm^{-2}$ and a $V_o$ of ~0.5 V *vs*. RHE in a pH 5 solution (Figure 8b4, see Table 1, Ref. 63). Notably, $V_o$ approached the values of a solid state P3HT:PCBM-based OSC, suggesting that electrons could be successfully extracted through the $TiO_2$ layer.[63] The photocurrent density increased with the $TiO_2$ thickness from 10 to 100 nm, achieving stability for $TiO_2$ layers of 75-100 nm (Figure 8, panel b4). Cyclic voltammetry analysis in the dark on catalyst-free devices (**Figure 14**a) evidenced that the reduction–oxidation processes of the CuI film were totally suppressed by 100 nm-thick $TiO_2$ layers, thus proving that the ALD-$TiO_2$ is capable to protect the underlying photoactive structures from the contact with the electrolyte. Contrariwise, thinner $TiO_2$ layers were able neither to provide adequate protection nor to form effective ESLs due to the irregular ALD growth onto the P3HT:PCBM surface, as a result of nucleation island formation[197] and Volmer-Weber type processes.[198] Stability tests indicated that $TiO_2$-protected photocathodes were stable for over 3 h operation at 0 V *vs*. RHE under simulated sunlight (Figure 14b).[63] Therefore, the $TiO_2$ layer, as obtained by low temperature ALD, provided an effective corrosion protection in both acidic and near-neutral pH as well as a stable and efficient ESL between the P3HT:PCBM and the EC.[63]

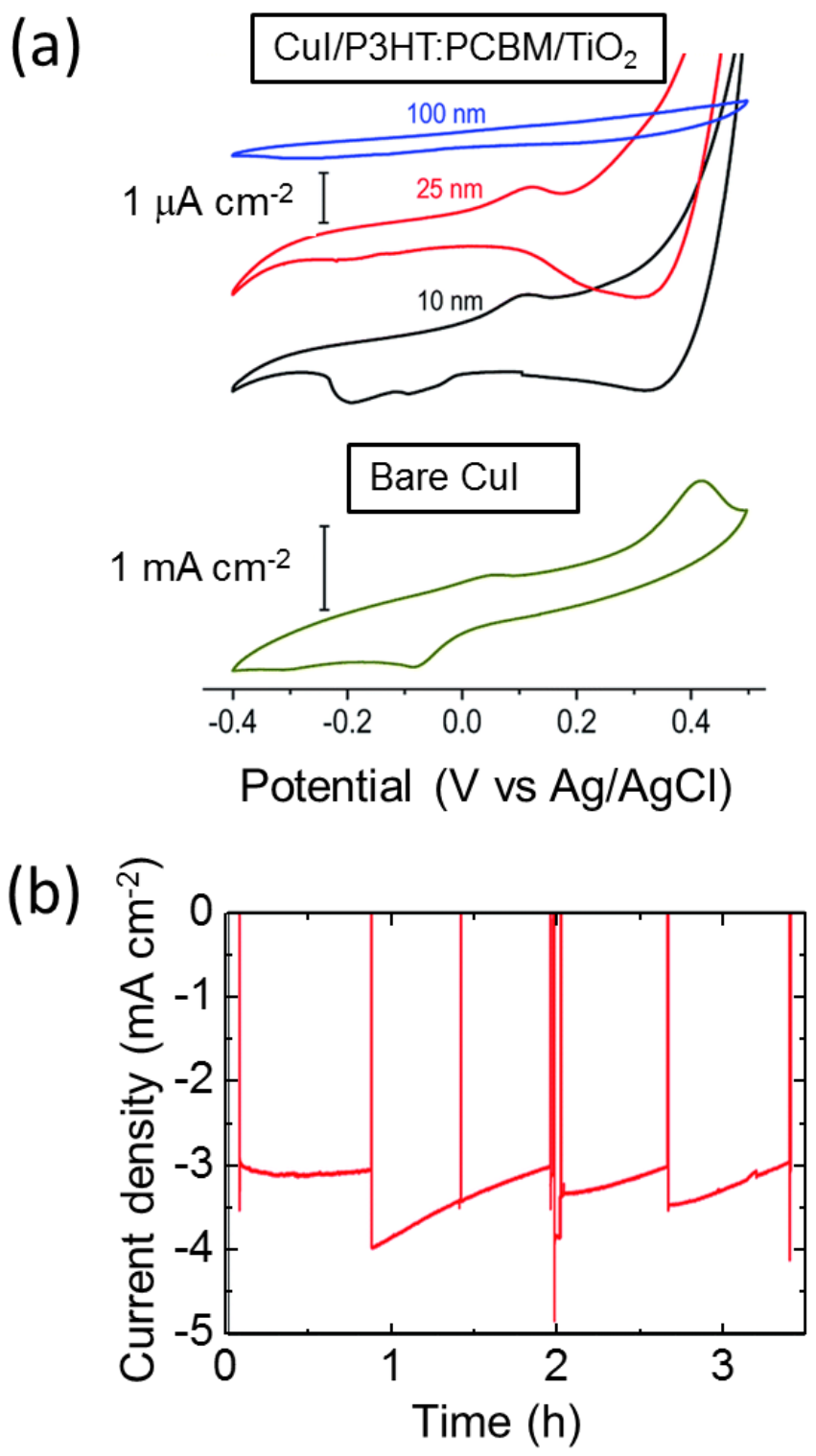

**Figure 14.** (a) Cyclic voltammetry measurements in the dark for three representative catalyst-free devices (FTO/CuI/P3HT:PCBM/$TiO_2$) using varied $TiO_2$ thicknesses (from 10 to 100 nm) in pH 5 electrolyte. The dark CV of a bare CuI sample is also shown for comparsion. (b) Stability test of the FTO/CuI/P3HT:PCBM/$TiO_2$(75 nm)/$RuO_x$ photocathode at 0 V *vs*. RHE. Adapted with permission.[63] Copyright 2017, The Royal Society of Chemistry.

Although both metallic Ti layers and compact ALD-$TiO_2$ proved to be effective for preserving the functionalities of the underlying photocathode structures, both the interfacial layers electronically separated the EC/electrolyte interface from the underlying structure (including the CSLs). Therefore, photocathodes implemented in this way should be better classified as PV-biased electrosynthetic junctions.[24] On one hand, this approach partially removes some of the PEC constraints, such as the rigorous alignment of the $E_{H+/H2}{}^{0}$ and the LUMO level of the photoactive material that is in direct contact with the electrolyte, as well as all the issues related to the electrochemical stability of the photoactive material that is in contact with aqueous solutions. On the other hand, however, the absence of any contact between the photoactive material and the electrolyte precludes any further exploitation of potentially interesting photoelectrode designs, such as an adaptive semiconductor/EC/electrolyte junction[92], thus hindering possible advantages of the carbon-based photocathodes [42-45,56,58,62,93] over their inorganic counterparts, which usually require the use of metal/metal oxide protective layers. [13,16,19,20a22,23]

### 5.3. Proton-conductive overlayers

In the previous sections, we analysed several examples of P3HT:PCBM-based photocathodes whose durability was negatively affected by EC detachment/dissolution and/or delamination/disruption of the photocathode structure.[54, 58,60,62,93] In order to overcome these issues, proton conducting overlays have been deposited by solution-processed techniques onto the photocathodes.[60,61,93] These materials should be water permeable and electrochemically stable in aqueous solution, in order to maintain the contact between the EC and the electrolyte and to allow electron transfer towards the electrolyte to be efficient. As an example, branched PEI protective layer was implemented to prevent the EC from dislodging and to stabilize the overall performance of the FTO/CuI/P3HT:PCBM/$TiO_2$/Pt photocathode (previously shown in Figure 8, panel a2).[60] PEI was chosen because of its good adhesive and coating properties, as well as its hydrophilicity, proton affinity and chelating properties on both ions and metals.[199] The PEC responses before and after the stability tests are shown in **Figure 15**. Neither the $V_o$ nor the $J_{0V\ vs.\ RHE}$ of the photocathodes were negatively affected by the PEI overlay, which resulted in more durable CuI-based photocathodes, still able to provide a $J_{0V\ vs\ RHE}$ of 1.68 mA $cm^{-2}$ after 3 h of operation.[60]

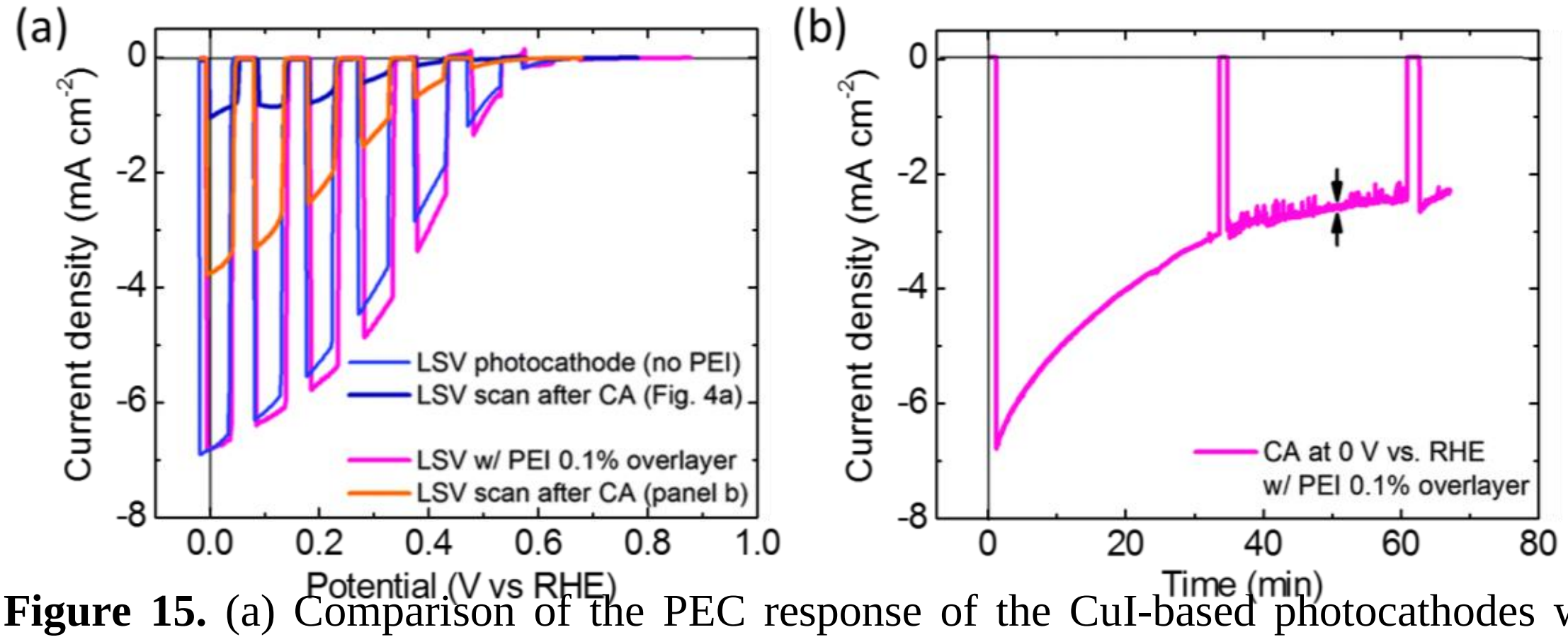

**Figure 15.** (a) Comparison of the PEC response of the CuI-based photocathodes with and without a PEI overlay in 0.5 M $H_2SO_4$ under simulated sunlight (AM1.5 light illumination, 100 W $cm^{-2}$), before and after stability tests. (b) Stability test of the CuI-based photocathode protected by the PEI overlay. Adapted with permission.[60] Copyright 2016, The Royal Society of Chemistry.

Recently, remarkable results for the stabilization of the photocathodes have been reached by adopting a solution-processed conductive and catalytic Pt on a carbon-tetrafluoroethylene-perfluoro-3,6-dioxa-4-methyl-7-octenesulfonic acid copolymer blend (Pt/C-Nafion) overlay within the FTO/graphene-based HSL/P3HT:PCBM/$TiO_2$ structures (Figure 12). The PEC response of the photocathodes that use GO as an HSL and Pt/C-Nafion as a protective electrocatalytic overlay has shown an improvement in PEC performance in pH 1 solution, reaching record high efficiencies ($\Phi_{saved,NPA,C}$ of 1.11% and a $\Phi_{saved,ideal}$ of 0.77% for FTO/GO/P3HT:PCBM/$TiO_2$/Pt/C-Nafion) for solution-processed P3HT:PCBM-based photocathodes (Figure 8, panel c2, see Table 1, Ref. 93). Noteworthy, these efficiency values approached the ones measured for P3HT:PCBM-based architectures produced through the evaporation of protective metallic Ti-based layers (based on PEDOT:PSS as HSLs, Figure 13a,b)[57] or PLD of $TiO_2$ (based on CuI as HSLs, Figure 8, panel a3)[60] (see Table 1, Ref. 57, 60). Moreover, the FTO/GO/P3HT:PCBM/$TiO_2$/Pt/C-Nafion photocathodes also exhibited remarkable durability (Figure 8c3), producing a $J_{0V\ vs\ RHE}$ of 4.14 mA $cm^{-2}$ after 1 h of continuous operation. Importantly, $J_{0V\ vs\ RHE}$ higher than 1 mA $cm^{-2}$ were measured even after 18 h operation. The durability was even higher for FTO/f-RGO/P3HT:PCBM/$TiO_2$/Pt/C-Nafion photocathodes, which had a $J_{0V\ vs\ RHE}$ higher than 1 mA $cm^{-2}$ after more than 20 h of operation. Thus, the use of both f-RGO as an HSL and a Pt/C-Nafion overlay enables a twofold increase in the photocathode durability in comparison with P3HT:PCBM-based photocathodes based on $WO_3$ as HSL,[58] tested up to 10 h (Figure 9b).

Overall, the use of a proton conductive overlay has to be considered a successful strategy to tackle the stability issues,[60,93] *e.g.*, EC detachment/dissolution and photocathode delamination effects.[54, 58,60,62,93]

## 6. Recent advances

This section will cover the most important advances recently achieved on carbon semiconductor-based photocathodes, following diverse, novel development routes. We consider: (1) the opportunity to tap into the large OSC materials catalog, testing novel materials substituting the P3HT:PCBM BHJ; (2) the capability to operate under different electrolyte pH; (3) the possibility to fabricate flexible and scalable devices.

### 6.1. Novel carbon-based photoelectrode materials: small molecules and oligomers

Based on the success of P3HT:PCBM-based photocathodes,[50,52-63,93] it is expected that many other OSC materials can serve as efficient photoactive materials in the realization of carbon-based photocathodes. Towards this direction, phthalocyanine-related small molecules and thiophene-based oligomers have recently raised interest.[76] Specifically, boron subphthalocyanine (SubPc) and subnaphthalocyanine (SubNc) chloride were used as acceptor materials, while alpha-sexithiophene (α-6T) was used as the donor material.[200] **Figure 16**a reports the chemical structures of these materials. Notably, sub(na)phthalocyanine derivatives are a class of phthalocyanine-related molecules made of three, rather than four, diiminoisoindoline units that are arranged around a boron atom. Based on their rather unique spectral and electronic features, such as nonplanar aromaticity,[201] they have been used as acceptor materials in OSCs[201b,202] in combination with α-6T small molecule as the donor.[200] Their capability to efficiently generate and collect charges under solar illumination is comparable or superior to that expressed by conventional fullerene-based acceptors.[200,203] In fact, although the fullerenes display a high electron mobility and large exciton diffusion length,[204] their reduced light absorption within the solar spectrum limits the maximum obtainable open-circuit voltage in OSCs.[205] Based on the knowledge acquired on OSCs,[206] bilayer planar hetrojunctions (2-PHJ) (*i.e.* ITO/PEDOT:PSS/D/A/BCP/C60/LiF/Al, in which D is the electron donor - α-6T, and A is the electron acceptor -SubPc or SubNc), as well as three-layer planar heterojunctions (3-PHJ) and hybrid heterojunction (HHJ) (including an additional layer between the D and the A moieties: SubNc or a blend of α-6T and SubNc) were investigated for photocathode architectures (Figure 16b). PEDOT:PSS was used as an HSL. 2,9-dimethyl-4,7-diphenyl-1,10-phenantroline (BCP) was used as an exciton blocking layer, while a $C_{60}$ interlayer was incorporated between the BCP and the EC (*e.g.*, $MoS_3$). Figure 16c shows the multilayer cascade energy level diagram of the most promising architecture, *i.e.* the ITO/PEDOT:PSS/α-6T/SubNc/BCP/$C_{60}$/$MoS_3$ photocathode. This photocathode has a $J_{0V\ vs\ RHE}$ of 3.6 mA $cm^{-2}$ and a $V_o$ of 0.69 V *vs*. RHE (Figure 16d). The HER-overpotential of the $MoS_3$, which was deposited directly onto ITO, was 0.15 V.[185] Therefore, the anodic shift of the HER onset potential (~0.84 V) approached the open-circuit voltage of the corresponding OSC using LiF/Al as a cathode. Despite the remarkable FoM valus achieved, the photocathode has shown a significant $J_{0V\ vs\ RHE}$ loss (~30%) after 5 min of endurance test. The decrease in the $J_{0V\ vs\ RHE}$ was not accompanied by a decrease in the $V_o$, suggesting that the photoactive core (*i.e.* α-6T/SubNc) was still intact, similar to what observed for P3HT:PCBM.[54,60,90] The decrease in the $J_{0V\ vs\ RHE}$ was ascribed to the degradation of the $C_{60}$ interfacial layer, possibly due to water diffusion and its reduction to a dianion state.[207]

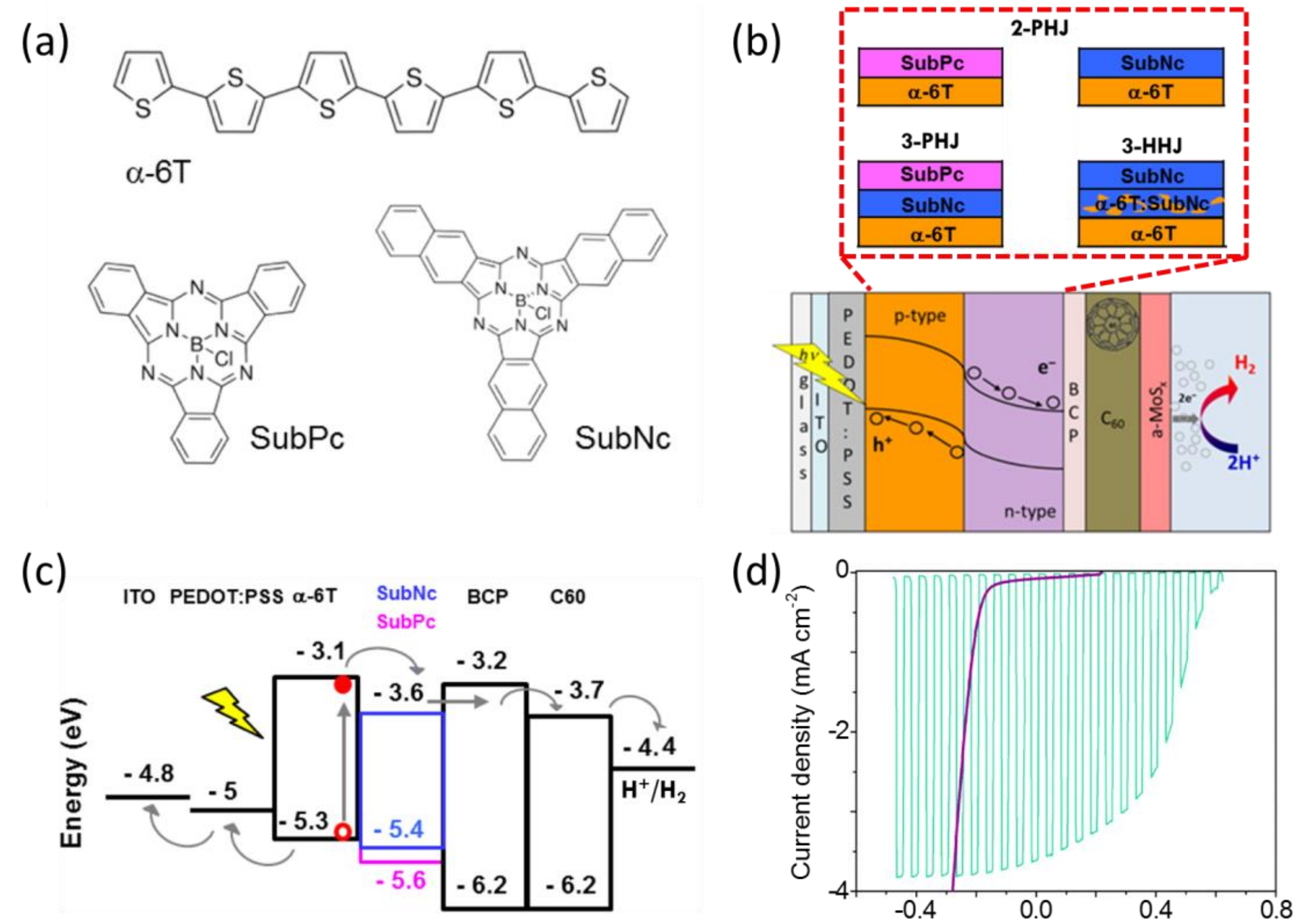

**Figure 16.** (a) Molecular structures of the α-6T, SubPc and SubNc. (b) Schematic representation of the 2-PHJ-, 3-PHJ, 3-HHJ-based photocathodes. (c) Energy band edge positions of the materials assembled in the 2-PHJ-, 3-PHJ, 3-HHJ-based photocathodes. (d) Photoelectrochemical response of the ITO/PEDOT:PSS/α-6T/SubNc/BCP/$C_{60}$/$MoS_x$ photocathode in 0.5 M $H_2SO_4$ solution under simulated sunlight (AM1.5 light illumination, 100 W $cm^{-2}$). Adapted with permission.[76] Copyright 2017, IOP Publishing.

Overall, these results preliminarly proved conceptually the possibility to extend the operating principles originally developed for P3HT:PCBM-based photocathodes to other OSC materials. Having considered that the efficiency of P3HT:PCBM-based OCSs (typically inferior to 5%)[162] is now largely outperformed by other organic blend formulations[208] (*e.g.*, efficiencies over 12% have been reached by using poly[(2,6-(4,8-bis(5-(2-ethylhexyl)thiophen-2-yl)benzo[1,2-b:4,5-b′]dithiophene)-co-(1,3-di(5-thiophene-2-yl)-5,7-bis(2-ethylhexyl)benzo[1,2-c:4,5-c′]dithiophene-4,8-dione)] –PBDB-T– as a donor, along with a small molecule acceptor)[208]), these results suggest that the use of other efficient OSC materials with intense and broad absorption, appropriate energy levels, and suitable crystallinity could lead to a rapid improvement in the performance of the photocathodes.

### 6.2. Widening pH operating window

The possibility to design a photocathode that is able to operate in a larger pH window is beneficial for the development of tandem architectures operating with neutral or alkaline solutions.[64, 209] Under these conditions, the photoanodes (having complementary electrochemical properties) of the tandem architecture usually exhibit a lower overpotential loss for the OER with respect to that exhibited in acidic solutions.[210] Notably, organic photoanodes for an OER based on SPs have recently been reported.[85] Although the photoanode sustained the OER over a wide pH range (from 1.9 to 12), the OER-related photocurrent density was found to increase with the pH. In fact, these results, together with the possibility to fabricate both photocathodes and photoanodes working under the same pH conditions, may lead to the development of all-organic tandem PEC water splitting cells. Furthermore, the possibility to operate under near-neutral pH aqueous conditions is of utmost interest because it allows sea and river water to be used, as easily available and non-hazardous/corrosive electrolyte.[211] This also releases the stability constraints of practical photoactive and catalyst components.[212] So far, the operation of the most efficient photocathode architectures has been demonstrated in most cases under acidic conditions.[47,52-54,56-62,76] The use of compact $TiO_2$, obtained by ALD in the P3HT:PCBM surface, enabled the P3HT:PCBM-based photocathode to exhibit a similar performance in both pH 5 and pH 1.36 solutions (see Table 1, Ref. 63). However, only recently the capability to operate in neutral or alkaline electrolytes started to be more extensively considered.[55,93] For example, Al-doped ZnO (AZO) nanocrystals[213] were exploited as an ESLfor fabricating P3HT:PCBM-based photocathodes (adopting insoluble cross-linked PEDOT:PSS as HSL and C/Pt as EC) working under neutral pH conditions.[55] The AZO-based photocathode has shown quite a poor performance at pH 2 and 5, due to the well known instability of the ZnO under acidic conditions, as evidenced by the Pourbaix diagram.[214] On the contrary, at pH 7, the photocathodes achieved a $J_{0V\ vs\ RHE}$ of 1.2 mA cm$^{-2}$ and a photocurrent density above 3 mA cm$^{-2}$ at -0.5 $V_{RHE}$. Unfortunately, after 60 min, the photocurrent density decreased by 50 %. X-ray photoelectron spectroscopy and Auger spectroscopy surface characterization on pristine and aged devices (after 1 h of operation at pH 7) revealed that metallic Zn was formed during the stability test, clearly indicating a partial reduction of the metallic oxide upon operation (*i.e.* $Zn^{2+} + 2e^{-} \rightarrow Zn^{0}$).

Recently, pH-universal P3HT:PCBM-based photocathodes adopting graphene derivatives as HSLs and a Pt/C-Nafion protective electrocatalytic overlay have been reported to operate under pH-universal conditions, ranging from acidic to alkaline electrolytes (**Figure 17**a,b).[93] In particular, $J_{0V\ vs\ RHE}$ of 1.64 (0.89), 1.51 (0.91), 1.41 (0.45) mA $cm^{-2}$ were measured for GO- (f-RGO-) based photocathodes at pH 4, 7 and 10, respectively. Currently, these architectures are the most efficient P3HT:PCBM-based photocathodes under neutral and alkaline conditions (see Table 1).[93] Figure 17c,d reports the stability tests at different pH values for the GO- and f-RGO- photocathodes.[93] In particular, after 5 h of continuous operation, GO- (f-RGO)-based photocathodes have shown a retention of the initial $J_{0V\ vs\ RHE}$ of 30% (64%) and 50% (66%) for pH 1 and 4, respectively.[93] After 20 h, they still displayed a retention of the initial $J_{0V\ vs\ RHE}$ of 12% (38%) and 27% (57%) at pH 1 and 4, respectively.[93] At pH 7 and 10 the photocurrent densities decreased rapidly after 5 h of operation.[93] This degradation was attributed to the electrochemical instability of the Pt/C-Nafion overlay in neutral and basic electrolytes.[215,216] In fact, the Pt dissolution/re-deposition mechanism or 3D Ostwald ripening[215] of the Pt/C-Nafion, due to the corrosion of C and Pt,[150a,216] could change the adhesion of the materials between the Pt/C-Nafion overlay. After the detachment/dissolution of the Pt/C-Nafion overlay, the underlying structure remained unprotected and exposed to the electrolyte, and the $H_2$ bubbling during the HER progressively degraded the photocathode surfaces.[93]

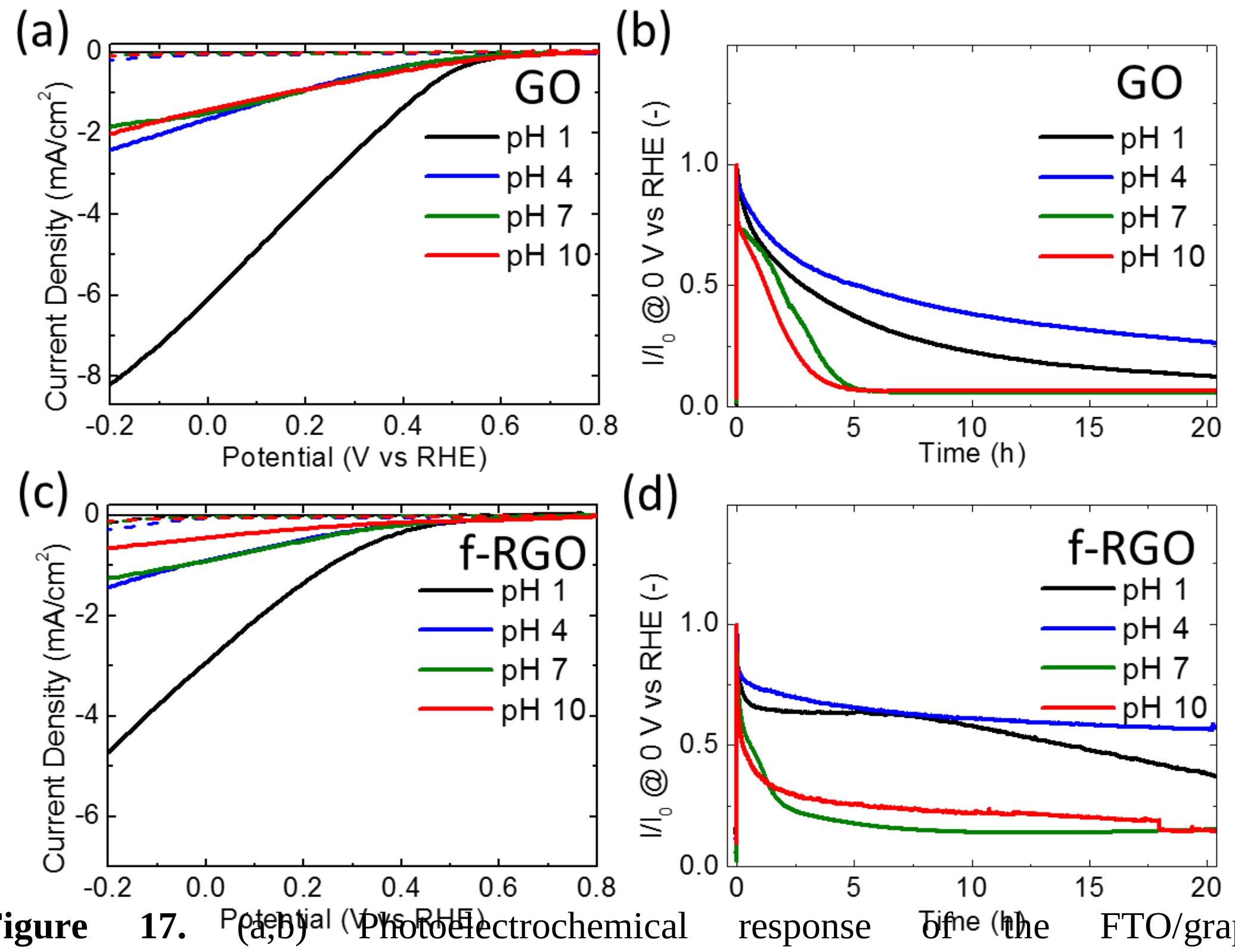

**Figure 17.** (a,b) Photoelectrochemical response of the FTO/graphene-based HSL/P3HT:PCBM/$TiO_2$/Pt/C-Nafion photocathode adopting GO (panel a) and f-RGO (panel b) as HSL under simulated sunlight (AM1.5 light illumination, 100 W $cm^{-2}$) in solution at different pH (1, 4, 7 and 10). (c,d) Stability test of the protected FTO/graphene-based HSL/P3HT:PCBM/$TiO_2$/Pt/C-Nafion adopting GO (panel c) and f-RGO (panel d) as HSL under simulated sunlight. Reprinted with permission.[93] Copyright 2017, American Chemical Society.

The results shown in this section clearly highlight the versatility of P3HT:PCBM-based photocathodes. In principle, they may operate in a wide range of electrochemical conditions, fulfilling the technological requirements of practical water splitting devices.

### 6.3. Solution processed, flexible and large-area architectures

Photoelectrodes based on carbon semiconductors could offer a low cost and high volume manufacturing thanks to their fast, solution-processed deposition onto flexible plastic substrates at low temperature.[69,98] Moreover, solution-processed fabrication can be intrinsically compatible with large-area photocathode fabrication.[69,98] Several all solution-processed architectures have been reported (see Table 1).[56,61,62,93] The most promising examples comprise the use of a dispersion/solution for the deposition of CSLs and ECs (PEDOT:PSS,[56] TMOs (*e.g.* $MoO_3$, NiO),[56] CuI,[61] $MoS_2$[62] and graphene derivatives[56,93] as HSLs; $TiO_2$ as an ESL;[61,62,93] $MoS_3$,[56,61] Pt[61,62,93] or C/Pt[61,93] as an EC). Notably, the architecture with TMOs did not include an ESL.[56] The first all solution-processed architecture adopting both an HSL and an ESL was FTO/CuI/P3HT:PCBM/$TiO_2$/catalyst/PEI.[61] **Figure 18**a reports a cross-sectional SEM image of a representative all solution-processed CuI-based photocathode. A key point in the successful realization of this architecture relates to the deposition, by solution-processed methods, of anatase $TiO_2$ nanoparticles, as obtained by a low temperature sol-gel method in aqueous media,[217] onto the P3HT:PCBM surface. In fact, the hydrophobic P3HT:PCBM film was treated with an $O_2$ plasma treatment in order to avoid wettability issues during the $TiO_2$ deposition from an aqueous dispersion. This allowed the formation of a homogeneous $TiO_2$ layer (100–120 nm thick, Figure 18a). A second $O_2$ plasma treatment of the $TiO_2$ surface was also applied in order to obtain a more homogeneous EC deposition. A thin PEI layer was finally deposited onto the photocathodes as protective coating (see Section 5.3).[60] The PEC performance of this all solution-processed device based on water dispersions of a Pt nanopowder or commercial Platinum on graphitized-carbon (*i.e.* Platinum on Vulcan – VC-Pt –),[218] or $MoS_3$ was slightly different (Figure 18b). In particular, the photocathode using Pt and $MoS_3$ achieved a $\Phi_{saved,ideal}$ of 0.67% and 0.38%, respectively (see Table 1, Ref. 61). The different values were attributed to the different HER-overpotential of the ECs.[185]

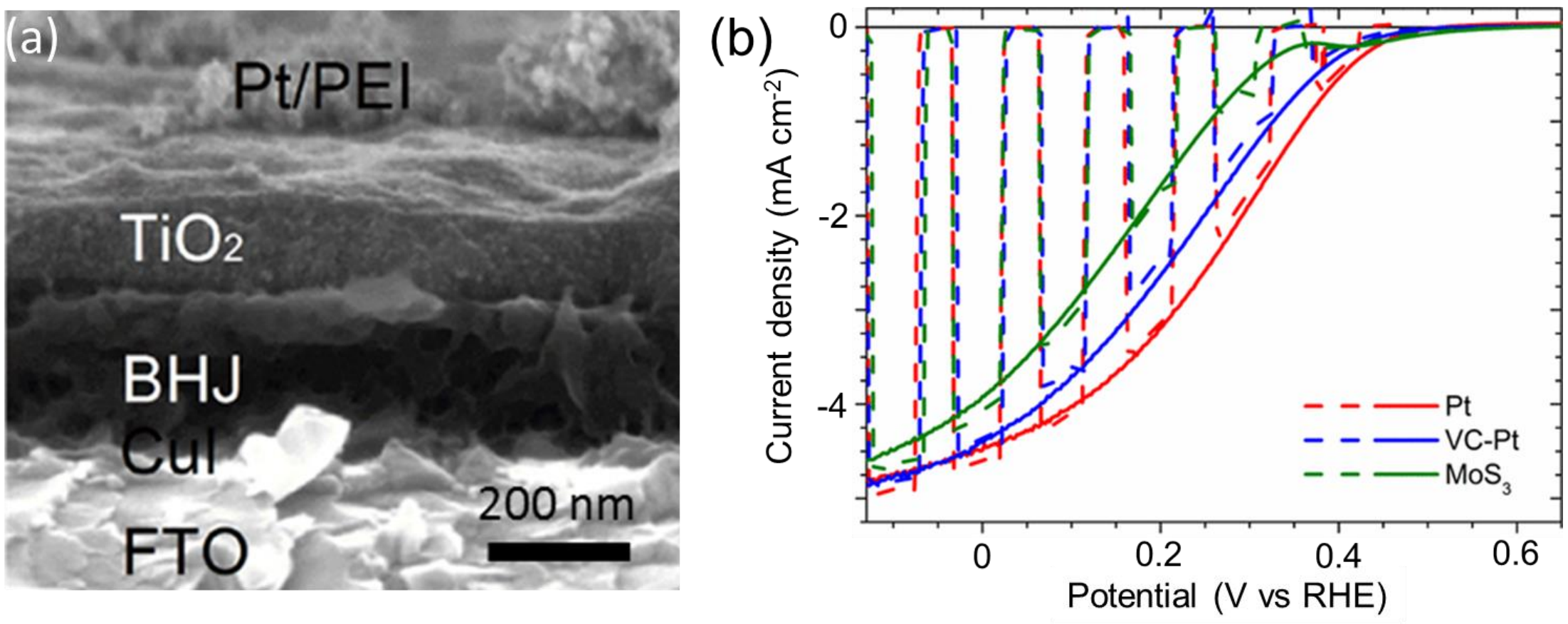

**Figure 18.** (a) Cross sectional SEM images of a representative all solution processed photocathode adopting the FTO/CuI/P3HT:PCBM/$TiO_2$/Pt/PEI architecture. (b) Photoelectrochemical response of the all solution-processed FTO/CuI/P3HT:PCBM/$TiO_2$/EC photocathode in pH 1 solution under simulated sunlight (AM 1.5 G light illumination, 100 W $cm^{-2}$) with different EC, *i.e.*: Pt, VC-Pt and $MoS_3$. Adapted with permission.[61] Copyright 2017, American Chemical Society.

Notably, photocathodes using 2D material as HSLs, *i.e.* $MoS_2$[62] and graphene derivatives,[93] have also been fabricated through all solution-based processing. In fact, as previously discussed in Section 5.1.4, the possibility to produce 2D materials from the exfoliation of their bulk counterpart in liquid[94] allowed the formulation of functional inks.[96] These inks can be then deposited onto different substrates by spin-coating protocols,[98] in a similar way to those adopted in OSCs. The up-scaling feasibility of the graphene-based devices was then demonstrated by fabricating a flexible 9 $cm^2$-area photocathode (Figure 19a,b). This achieved a $J_{0V\ vs\ RHE}$ of 2.80 mA $cm^{-2}$, a $V_o$ = 0.45 V *vs.* RHE, a $\Phi_{saved,NPA,C}$ = 0.31%, and a $\Phi_{saved,ideal}$ = 0.23% (Figure 19c). For the 9 $cm^2$-area device, the poor performance (in comparison to the 1 $cm^2$-area one) was attributed to the series resistance of the photocathodes. In fact, the $R_s$ values observed for the ITO-PET were higher than those of the FTO, which caused the decrease in the $V_{mpp}$ for the 9 $cm^2$-area photocathode (0.26 V *vs.* RHE) with respect to that of 1 $cm^2$-area one (0.17 V *vs.* RHE). This caused a decrease in the FF from 0.21 in the 1 $cm^2$-area configuration to 0.16 in the 9 $cm^2$-area configuration, as illustrated in Figure 19d.

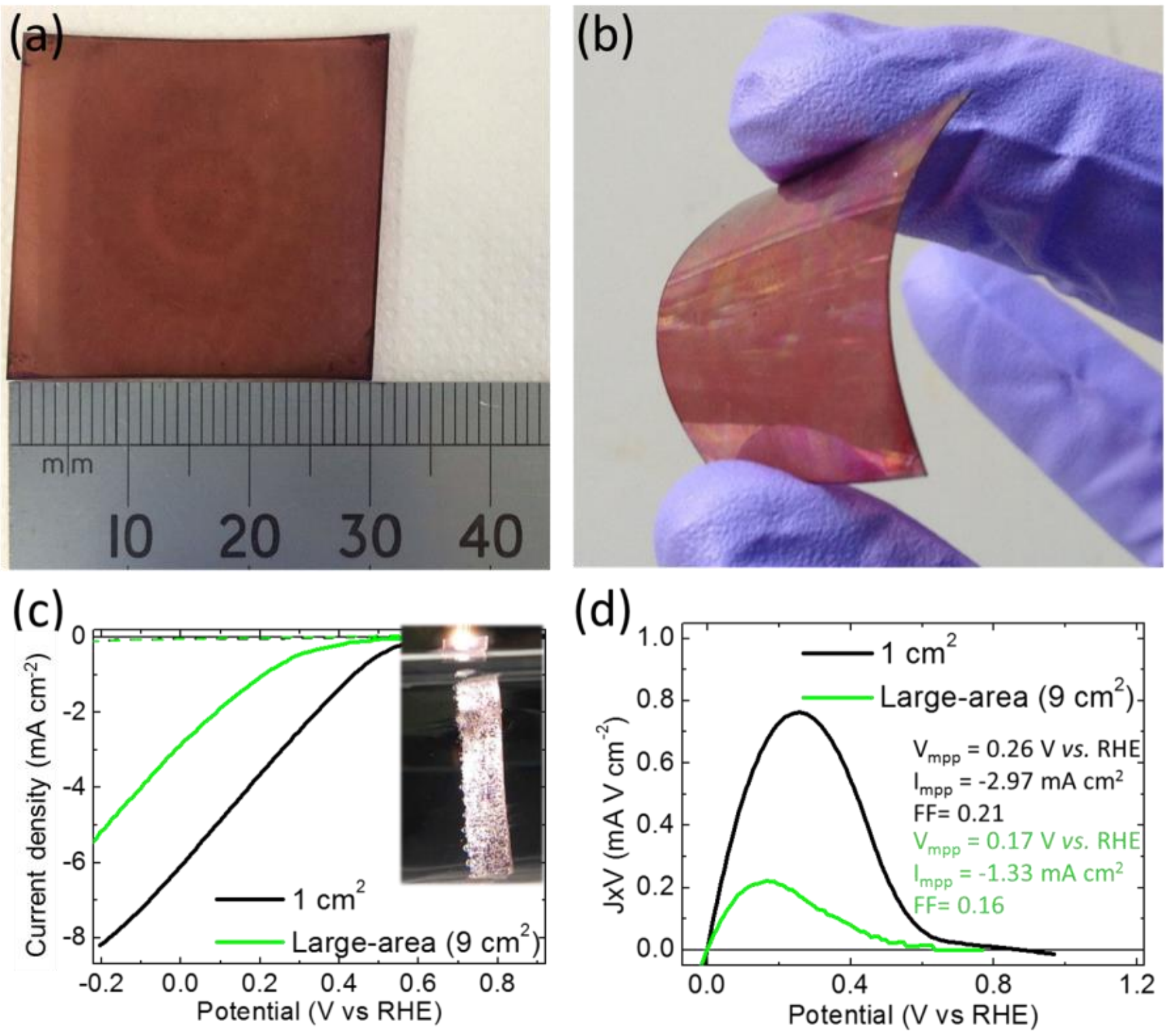

**Figure 19.** (a,b) Photographs of a representative solution-processed large-area (9 cm$^2$) ITO-PET/GO/rr-P3HT:PCBM/$TiO_2$/Pt/C-Nafion photocathode before (panel a) and after bending (panel b). (c) Photoelectrochemical response and (d) current density × potential *vs.* potential curves that were measured for the ITO-PET/GO/rr-P3HT:PCBM/$TiO_2$/Pt/C-Nafion photocathode, with an area of 1 cm$^2$ and 9 cm$^2$ (black and olive lines, respectively) in pH 1 solution under dark (dashed lines) and simulated sunlight (solid lines). The values of the $V_{mpp}$, $J_{mpp}$ and FF of the photocathodes are also reported in panel (d), showing a decrease in the FF by increasing the area of the photocathode. Adapted with permission.[93] Copyright 2017, American Chemical Society.

The $R_s$ of both FTO and ITO-PET substrates can be reduced by integrating metal grids onto ITO or FTO (*e.g.*, electroplated Cu grids)[219] or by connecting the ITO or FTO through holes to a backside metallic electrode.[220] The results demonstrated that it is possible to realize large-area P3HT:PCBM photocathodes by uniformly depositing the different materials using scalable solution-processed techniques.

### 6.4 Novel 3D nanostructured architectures

Practical photocathode designs need to simultaneously maximize the light absorption and the extraction of the photogenerated charges.[221] In the particular case of P3HT:PCBM-based photocathodes, the trade-off between the minimal P3HT:PCBM layer thickness, necessary to achieve substantial light absorption ($L_{abs}$≈200 nm at 450 nm for P3HT film), and the diffusion length of photogenerated excitons and charge carriers within the P3HT film ($L_{exc}$≈5 nm)[222] limits the efficiencies of the photocathodes with a simple stacked-layer architecture. The direct nanostructuring of the absorber, successfully reported for many materials of interest in water-splitting,[223] is beneficial for increasing the surface area at the semiconductor-liquid interface,[45] as well as for improving the light trapping. Unfortunately, it lacks the orthogonalization of light absorption and the carrier collections, which are necessary in the case of absorbers with poor charge transport properties.[224] In fact, in order to achieve absorption/collection decoupling, the HSL can be nanostructured,[225] creating an high interfacial area, if compared to the flat case, while preserving optical transparency.[59] Subsequently, a nm-thick structure of a photoactive layer can be deposited onto the large specific area of the HSL. This architecture resembles the one of host scaffold/guest absorber architectures that were developed for dye-sensitized solar cells[226] and $O_2$ evolution $Fe_2O_3$-based photoanodes.[227] Recently, this approach has been applied to hybrid organic/inorganic photocathode architectures by introducing a nanostructured $MoO_3$ scaffold as the HSL,[59] thus combining the intrinsic $MoO_3$ properties (*e.g*. transparency,[228] energetics[229] and charge transport[230]), with the structural features of nanostructured materials (*e.g*. enhanced light scattering[231] and increased surface area[221,232]). In particular, the morphological character of $MoO_3$ films was controlled by operating the PLD process at different $O_2$ backgrounds.[233] The low gas pressure (5 Pa) produced compact $MoO_3$ films, while pressures higher than 10 Pa produced vertically aligned $MoO_3$ *lamellae* (**Figure 20**a).[59] Then, various P3HT:PCBM-based photocathodes were fabricated using compact $MoO_3$ (compact-$MoO_3$) or nanostructured $MoO_3$ (ns-$MoO_3$) as an HSL and nm-thick P3HT:PCBM layers (~30 nm) (Figure 20b,c).[59] Light trapping phenomena were observed moving from compact-$MoO_3$ to nanostructured-$MoO_3$ (Figure 20d),[59] as a consequence of the Rayleigh/Mie-type interaction.[234] Overall, ns-$MoO_3$ has shown a decrease in transparency with respect to the compact-$MoO_3$, and the diffuse transmittances of the different ns-$MoO_3$ were almost superimposed upon the total transmittances. This means that the light is scattered from the ns-$MoO_3$ towards the photoactive material in the photocathode architecture,[59] enabling the P3HT:PCBM layer to absorb more than 95 % of the incoming light in the 400-600 nm range. Consequently, the ns-$Mo_3$-based photocathodes have shown an increase in the photocurrent density when the PLD pressure was increased during deposition of $MoO_3$ film (Figure 20e).[59] The ns-$MoO_3$-based architecture with the best performance exhibited a photocurrent of 1.3 mA $cm^{-2}$ at 0.18 V *vs.* RHE and a $\phi_{saved,ideal}$ of 0.37 % (see Table 1, Ref. 59), much higher than the value exhibited by compact-$MoO_3$ (50 μA $cm^{-2}$). The photocathodes displayed the same $V_o$ values around 0.65 V *vs.* RHE, which excludes difference of energy levels between the materials assembling the photocathodes,[59] while ascribes the enhancement of the PEC performance to the $MoO_3$ nanostructurization-induced light trapping.[59] Overall, these results proved that the out-of-plane HSL nanostructuring effectively decoupled the $L_{abs}$ from the $L_{exc}$, thus enabling an enhancement in the PEC performance of the P3HT:PCBM photocathodes.

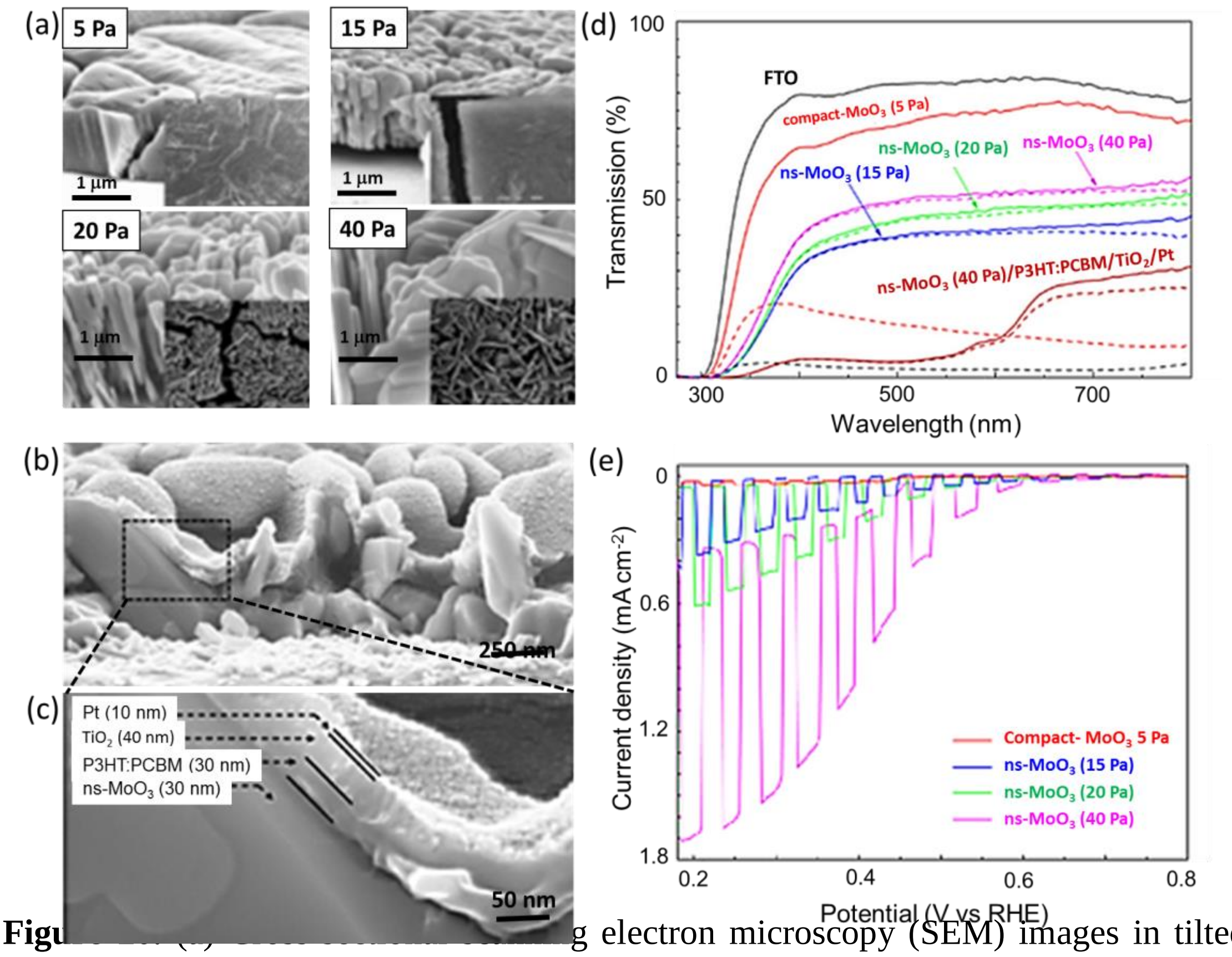

**Fig**[illegible]g electron microscopy (SEM) images in tilted cross-sectional view (main) and top view (inset) of $MoO_3$ films deposited through PLD at different $O_2$ pressures: 5 Pa, 15 Pa, 20 Pa and 40 Pa. The images highlight the structural and morphological dependence of the film on the deposition conditions. (b) Cross-sectional SEM images of the complete host/guest hybrid organic/inorganic photocathode architecture, employing a ns-$MoO_3$ deposited through PLD at an $O_2$ gas pressure of 40 Pa. (c) SEM magnification of the image showed in panel (b), demonstrating the different photocathode layers: the ns-$MoO_3$ (500 nm) as an HSL; the P3HT:PCBM (30 nm) as a photactive layer; the $TiO_2$ (40 nm) as an ESL; the Pt (10 nm) as an EC. (d) UV/Vis/nIR total transmittance (solid) and diffuse transmittance (dashed) spectra of FTO-coated glass substrate (black), compact-$MoO_3$ (red) and three different morphologies of ns-$MoO_3$ deposited at 15 Pa (blue), 20 Pa (green) and 40 Pa (pink) of $O_2$ pressure. For the ns-$MoO_3$ deposited at 40 Pa, the transmittance of the full photocathode architecture is also shown (wine-colored). (e) Photoelectrochemical response of the photocathodes in $H_2SO_4$:$Na_2SO_4$ electrolyte at pH 1.37 under simulated sunlight (AM 1.5 G light illumination, 100 W $cm^{-2}$) with compact and ns-$MoO_3$ as HSLs. Adapted with permission.[59] Copyright 2018, John Wiley and Sons.

## 7. Outlook

In this review article, we discussed the realization of $H_2$-evolving carbon semiconductor-based photocathodes, providing an overview of the development of the photocathode architectures currently developed. In particular, we focused our critical analysis on the up-to-date research efforts towards the engineering of the charge-selective layers and the enhancement of the stability. The possibility to formulate the photocathode materials that comprise inorganic charge-selective layers and electrocatalysts in a liquid phase already allowed for a straightforward fabrication of large-area devices without any challenging or expensive fabrication processes. However, both the screening of the carbon semiconductors as photoactive materials and the design of optimal photoelectrode architectures (*e.g.* the use of viable noble-metal free HER-ECs) are still in early stages of development. In particular, major results on carbon-based photocathodes have been achieved by focusing only on the archetypical bulk heterojunction established in organic photovoltaics, *i.e.* P3HT:PCBM, whose efficiency in such devices (typically inferior to 5%) is now largely outperformed by other organic blend formulations. For example, efficiencies over 12% have been reached by using PBDB-T as a donor, and small molecule acceptor.[208] A few works have already demonstrated the use of organic molecules and oligomers as photocathode materials.[46,76,207, 235 ] Simultaneously with the writing of this work, novel carbon-based photocathodes based on poly[N-9'-heptadecanyl-2,7-carbazole-alt-5,5-(4',7'-di-2-thienyl-2',1',3'-benzothiadiazole)] (PCDTBT),[236] polyterthiophenes (PTTh)[209b] and conjugated 2D covalent organic frameworks have been reported.[237]

Furthermore, it is crucial, though challenging, that photocathode materials are electrochemically stable in aqueous electrolytes in order to implement them into practical devices. So far, these properties have been verified mainly only for thiophene-based polymers,[90] including P3HT[90] and a few other low-band gap polymers.[90b] Notably, overcoming the stability issues of the photocathode materials is a major advantage compared to the state-of-the-art technologies,[12,13] in which the need to implement expensive protective strategies typically hindered the commercialization of monolithic water splitting devices.[22c,91] Moreover, it is also important that, in the presence of unstable absorbers, the incorporation of protective layers avoids absorber/electrolyte contact, thus the photo electrochemical architectures move towards photovoltaic-buried junctions.[24] This means that the advantages that could possibly arise from the strong short-spatial range electrochemical coupling between the photoactive material and the electrolyte are excluded from the outset.[92] It is worth noting that photovoltaic-buried junctions were also reported here because of the implementation of an interfacial compact layer as a stabilizing strategy.[57,63] However, in the presented case, their role was initially to avoid the degradation of the charge-selective layers, and not to protect the organic photoactive layer.

The results here discussed highlight that the great challenges and opportunities for the use of carbon semiconductors in the field of $H_2$-evolving photocathodes as well as for electrochemical applications in general.

## List of abbreviations

PEC: photoelectrochemical; HER: hydrogen evolution reaction; OER: oxygen evolution reaction; FoM: Figure of Merit; SHE: standard hydrogen electrode; ALD: atomic layer deposition; PV: photovoltaic; SP: semiconducting polymer; OSC: organic solar cell; LUMO: lowest unoccupied molecular orbital; BHJ: bulk heterojunction; 3-PHJ: three-layer planar heterojunctions; ORR: oxygen reduction reaction; RHE: reversible hydrogen electrode; EC: electrocatalyst; TCO: transparent conductive oxide; ITO: Indium Tin Oxide; FTO: Fluorine doped Tin OixdeCSL: charge selective layer; HSL: hole selective layer; ESL: electron selective layer; EC: electrocatalyst; HER-EC: electrocatalyst for HER; SEM: scanning electron microscopy; PLD: pulsed layer deposition; TMO: transition metal oxide; ECP: electrically conductive polymer; TMD: transition metal dichalcogenide; $\eta_{STH}$: solar-to-hydrogen conversion efficiency; $J_{sc}$: the short-circuit photocurrent density; $\eta_F$: the Faradaic efficiency; AM1.5G: standard solar illumination (simulated sunlight); P3HT: regio-regular poly(3-hexylthiophene-2,5-diyl); PCBM: phenyl-C61-butyric acid methyl ester; α-6T: alpha-sexithiophene; PBDB-T: poly[(2,6-(4,8-bis(5-(2-ethylhexyl)thiophen-2-yl)benzo[1,2-b:4,5-b′ ]dithiophene)-co-(1,3-di(5-thiophene-2-yl)-5,7-bis(2-ethylhexyl)benzo[1,2-c:4,5-c′ ]dithiophene-4,8-dione)]; PEDOT: poly(3,4-ethylenedioxythiophene); PEDOT:PSS: poly(3,4ethylenedioxythiophene):poly(styrene sulfonate); PBTh: poly(2,2-bithiophene); PANI: polyaniline; PPY: Polypyrrole; SubPc: boron subphthalocyanine chloride; SubNc: subnaphthalocyanine chloride; BCP: 2,9-dimethyl-4,7-diphenyl-1,10-phenantroline; PEI: polyethyleneimine; Nafion: tetrafluoroethyleneperfluoro-3,6-dioxa-4-methyl-7-octenesulfonic acid copolymer; GO: graphene oxide; RGO: reduced graphene oxide; g-$C_3N_4$: graphitic carbon nitride; Fc/$Fc^+$: ferrocene/$ferrocene^+$ redox couple; BZQ/$BZQ^{\cdot-}$: benzoquinone/$benzoquinone^{\cdot-}$ redox couple; $J_{0V\ vs\ RHE}$: cathodic photocurrent density at 0 V *vs*. RHE; $V_o$: onset potential; FF: fill factor; $V_{mpp}$: maximum power point; $\Phi_{saved,NPA,C}$: ratiometric power-saved efficiency relative to a non-photoactive (NPA) dark electrode with an identical catalyst (C): $\Phi_{saved,ideal}$: ratiometric power-saved efficiency relative to an ideally non-polarizable RHE;

## Acknowledgements

The authors thank Leyla Najafi, Antonio Esau Del Rio Castillo, Alberto Ansaldo, Mirko Prato, Valentino Romano for useful discussions. The authors acknowledge funding from the European Union's Horizon 2020 research and innovation programme under grant agreement No. 696656 – GrapheneCore1.